\documentclass[a4paper,11pt]{article}
\pdfoutput=1
\usepackage{jheppub}

\usepackage{slashed}
\usepackage{bm,wasysym}
\usepackage[normalem]{ulem}

\usepackage{amsmath}
\usepackage{amssymb}
\usepackage{bbm}
\usepackage{feynmp}
\usepackage{graphicx}
\usepackage{slashed}
\usepackage{subfigure}
\usepackage[dvipsnames]{xcolor}

\newcommand{\wilson}[2][]{\mathcal{C}^{#1}_{#2}}
\newcommand{\op}[2][]{\mathcal{O}^{#1}_{#2}}
\newcommand{\bra}[1]{\big\langle{#1}\big\vert}
\newcommand{\ket}[1]{\big\vert{#1}\big\rangle}
\newcommand{\krf}{|\vec{k}_\text{$N\pi$-RF}|}
\newcommand{\qrf}{|\vec{q}_\text{2$\ell$-RF}|}
\renewcommand{\Im}[1]{\operatorname{Im}\left\lbrace{#1}\right\rbrace}
\renewcommand{\Re}[1]{{\rm Re}\left\lbrace{#1}\right\rbrace}

\DeclareMathOperator{\sign}{sgn}
\DeclareMathOperator{\argmax}{arg\ max}
\renewcommand{\[}{\big[}
\renewcommand{\]}{\big]}

\newcommand{\dd}{{\rm d}}
\newcommand{\refapp}[1]{appendix~\ref{sec:app:#1}}
\newcommand{\refeq}[1]{eq.~(\ref{eq:#1})}

\newcommand{\reftab}[1]{table~\ref{tab:#1}}

\newcommand{\refsec}[1]{section~\ref{sec:#1}}

\newcommand{\eps}{\varepsilon}
\newcommand{\para}{\parallel}

\newcommand{\gfermi}{G_\text{F}}
\newcommand{\mlamB}{m_{\Lambda_b}}
\newcommand{\mlam}{m_{\Lambda}}

\newcommand{\order}[1]{\mathcal{O}\left({#1}\right)}
\newcommand{\GeV}{\textrm{GeV}}

\newcommand\tabvsptop{\rule{0pt}{2.6ex}}
\newcommand\tabvspbot{\rule[-1.5ex]{0pt}{0pt}}

\numberwithin{equation}{section}
\allowdisplaybreaks

\def\slash#1{\setbox0=\hbox{$#1$}\dimen0=\wd0
      \setbox1=\hbox{/} \dimen1=\wd1 \ifdim\dimen0>\dimen1
      \rlap{\hbox to \dimen0{\hfil/\hfil}} #1                        \else
      \rlap{\hbox to \dimen1{\hfil$#1$\hfil}}
      /   \fi}

\newcommand{\lsim}{
\mathrel{\hbox{\rlap{\hbox{\lower4pt\hbox{$\sim$}}}\hbox{$<$}}}}

\newcommand{\gsim}{
\mathrel{\hbox{\rlap{\hbox{\lower4pt\hbox{$\sim$}}}\hbox{$>$}}}}

\allowdisplaybreaks[2]

\title{Angular Analysis of the Decay $\Lambda_b \to \Lambda (\to N \pi) \ell^+\ell^-$}
\author{Philipp B\"oer,}
\emailAdd{boeer@physik.uni-siegen.de}
\author{Thorsten Feldmann,}
\emailAdd{thorsten.feldmann@uni-siegen.de}
\author{Danny van Dyk}
\emailAdd{vandyk@physik.uni-siegen.de}
\affiliation{Theoretische Physik 1, Naturwissenschaftlich-Technische Fakult\"at,
Universit\"at Siegen,\newline Walter-Flex-Stra\ss{}e 3, D-57068 Siegen, Germany}

\abstract{We study the differential decay rate for the rare $\Lambda_b \to \Lambda (\to N \pi)\ell^+\ell^-$
   transition, including a determination of the complete angular distribution, assuming 
   unpolarized $\Lambda_b$ baryons.
   On the basis of a properly chosen parametrization of the various helicity amplitudes,
    we provide expressions for the angular observables within the Standard Model and a subset
    of new physics  models with chirality-flipped operators. 
    Hadronic effects at low recoil are estimated by combining information from lattice QCD
    with (improved)  form-factor relations in Heavy Quark Effective Theory. 
    Our estimates for large hadronic recoil -- at this stage -- are still rather uncertain because 
    the baryonic input functions are not so well known, and non-factorizable spectator effects 
    have not been worked out systematically so far.
    Still, our phenomenological analysis of decay asymmetries and angular observables
    for  $\Lambda_b \to \Lambda (\to N \pi)\ell^+\ell^-$ reveals that this decay mode 
    can provide new and complementary constraints on the Wilson coefficients in radiative
    and semileptonic $b \to s$ transitions compared to the corresponding mesonic modes.
}

\keywords{Rare Quark Decays, Heavy Quark Expansion, Soft-Collinear Effective Theory, Baryonic Form Factors}

\preprint{SI-HEP-2014-16, QFET-2014-11, EOS-2014-01}

\begin{document}

\maketitle

\section{Introduction}

Rare decays based on radiative or semi-leptonic $b \to s$ transitions 
offer various possibilities to test the predictions for flavour-changing
neutral currents (FCNCs) in the Standard Model (SM) against new physics (NP)
extensions (for comprehensive summaries of theoretical and experimental 
aspects, see e.g.\ \cite{Buchalla:2008jp,Antonelli:2009ws,Bediaga:2012py}).
In the past -- notably during the ``$B$-factory'' era \cite{Bevan:2014iga} -- 
the main phenomenological focus was on inclusive distributions 
($B \to X_s\gamma$ and $B \to X_s\ell^+\ell^-$) or exclusive decay observables
(e.g.\ in $B \to K^*\gamma$, $B \to K^{(*)}\ell^+\ell^-$,  \ldots )
for \emph{mesonic} decays.
With the recent $b$-physics program at LHC (and here, in particular,
the dedicated LHCb experiment) not only more precise measurements of 
radiative $B$- and $B_s$-meson decays,
but also information on baryonic modes 
like $\Lambda_b \to \Lambda\ell^+\ell^-$ with reasonable accuracy
will become available \cite{Aaij:2013hna}.
(For an incomplete list of
previous phenomenological studies, see e.g.\
\cite{Hiller:2001zj,Chen:2001zc,Aliev:2010uy,Azizi:2013eta,Gutsche:2013pp} and references therein).

Exclusive hadronic decays, by definition, are theoretically challenging
because the calculation of decay amplitudes induces a number 
of hadronic uncertainties related to long-distance QCD dynamics. 
For $b \to s\ell^+\ell^-$ transitions this includes hadronic transition 
form factors, which parametrize the ``naively'' factorizing contributions
from $b \to s\gamma$ and $b\to s\ell^+\ell^-$ operators.
In addition, systematic uncertainties related to 
non-factorizable effects appear, where the short-distance dynamics 
is induced by \emph{hadronic} $b \to s$ operators, while the 
radiation of the photon or charged lepton pair is linked 
to the long-distance hadronic transition.
Baryonic transitions, at first glance, seem to suffer
from even larger hadronic uncertainties than their mesonic
counterparts since transition form factors and
hadronic wave functions  are only poorly known,
and the analysis of the spectator dynamics 
is more complicated \cite{Wang:2011uv}.
Recent progress with respect to 
$\Lambda_b \to \Lambda$ form factors
includes lattice-QCD results in the heavy-mass limit
\cite{Detmold:2012vy} (valid 
at low and intermediate recoil), and a sum-rule analysis 
of spectator-scattering corrections to form factor relations 
at large recoil \cite{Feldmann:2011xf}. 
Also, a better theoretical understanding
of the $\Lambda_b$ wave function 
(in the form of light-cone distribution amplitudes)
has been achieved recently \cite{Ali:2012pn,Bell:2013tfa,Braun:2014npa}.

However, as we will also argue in this paper, exclusive
modes like $\Lambda_b \to \Lambda \ell^+\ell^-$ do provide
interesting phenomenological potential. 
Exploiting the full set of angular observables
that can be derived from the analysis of the subsequent 
$\Lambda \to N\pi$ decay, one may obtain
information on the underlying short-distance weak interactions 
that is complementary to the analogous mesonic decay observables
(see e.g. \cite{Kruger:2005ep,Altmannshofer:2008dz,Bobeth:2008ij,Egede:2008uy,Bobeth:2010wg,Becirevic:2011bp,Bobeth:2012vn}).
This is mainly a consequence of the fact that
the subsequent weak decay $\Lambda \to N\pi$ is parity violating,
while the strong decay $K^*\to K\pi$ in the mesonic counterpart case, 
$B \to K^*\ell^+\ell^-$, is not.\footnote{The background from direct $\Lambda_b\to N\pi\ell^+\ell^-$ decays
is expected to be low: First, the direct decay is relatively suppressed by $|V_{td}/V_{ts}|$. 
Second, the signal with an intermediate $\Lambda$ baryon 
can be distinguished by requiring a sizable displacement between the 
$\Lambda_b \ell^+\ell^-$ and $N \pi$ decay vertices.}
Furthermore, within such an analysis, independent information on
the hadronic parameters themselves can be extracted from 
experimental data. 
In particular, form-factor relations that arise in the infinite-mass
limit for the heavy $b$-quark, can be tested (notice that the
number of independent form factors for $\Lambda_b \to \Lambda$
transitions in that limit is smaller than for $B \to K^*$ transitions:
2 at low recoil \cite{Mannel:1997xy}, and one at large recoil \cite{Feldmann:2011xf,Mannel:2011xg}).

The outline of this paper is as follows.
In the next section, we will briefly summarize our notation and conventions
regarding the short-distance operator basis in the weak effective Hamiltonian.
In \refsec{decay} we derive the necessary expressions to describe the 
$\Lambda_b \to \Lambda(\to N\pi) \ell^+\ell^-$ differential decay rate.\footnote{%
    We will restrict ourselves to the case of unpolarized $\Lambda_b$ baryons,
    as the $\Lambda_b$ polarization in the LHCb setup has been measured 
    to be small \cite{Aaij:2013oxa}, and polarization effects in the symmetric ATLAS and
CMS detectors will average out.}
To this end, we define the relevant kinematic variables, define the 
general set of $\Lambda_b \to \Lambda$ form factors in the helicity basis,
and discuss the hadronic couplings appearing in the $\Lambda \to N\pi$ decay.
From this we derive expressions for the angular observables in terms of
transversity amplitudes.
Section \ref{sec:pheno} is dedicated to the phenomenological analysis 
of interesting observables. Specifically, 
we consider the fraction of transverse dilepton polarization,
the leptonic, baryonic and mixed forward-backward asymmetries,
and a number of certain ratios of angular observables where 
either short-distance effects or form-factor uncertainties drop
out to first approximation. We also work out the simplifications
that arise from the heavy-quark expansion at low or large recoil.
We conclude this section with numerical predictions
for a selection of observables, 
making use of the available theoretical and phenomenological
information on the relevant input parameters, and compare to
presently available experimental data in the low-recoil region.
We conclude with a summary and outlook. Some technical details 
are summarized in the appendices.

%%
%% Effective Hamiltonian
%%

\section{Effective Hamiltonian for $b\to s \ell^+\ell^-$ Transitions}

The effective weak Hamiltonian for
$b \to s \ell^+\ell^-$ transitions ($|\Delta B| = |\Delta S| = 1$)
in the SM (see e.g.\ \cite{Buchalla:1995vs,Chetyrkin:1996vx,Buras:2011we}) contains radiative ($b \to s\gamma$)
and semi-leptonic ($b \to s\ell^+\ell^-$)
operators as well as hadronic operators ($b\to sq\bar q$, $b\to sg$).
For the radiative and semileptonic operators, we find it convenient to use the normalization convention
\begin{equation}
    {\cal H}^\text{eff} \Big|_{\rm SM, naive}
    = \underbrace{\frac{4 G_{\rm F}}{\sqrt 2} \, V_{tb} V_{ts}^* \,
    \frac{\alpha_e}{4 \pi}}_{N_1} \, \sum_{i=7,9,10} \wilson{i}(\mu) \, \op{i}
    \,,
\end{equation}
with the SM operators
\begin{align}
\op{7(')}
    & = \frac{m_{b}}{e} \[\bar s \sigma^{\mu\nu} P_{R(L)} b\] F_{\mu\nu}\,,\quad
    \op{9}
     = \[\bar s \gamma^\mu P_{L} b\]\[\bar\ell \gamma_\mu \ell\]\,, \quad
    \op{10}
     = \[\bar s \gamma^\mu P_{L} b\]\[\bar\ell \gamma_\mu \gamma_5 \ell\]\,.
\end{align}
For simplicity, we only take into account the factorizable quark-loop contributions of the hadronic operators
$\op{1-6}$ and $\op[g]{8}$, which can be lumped into effective Wilson coefficients $\wilson{7}{}^{\rm eff}$ and $\wilson{9}{}^{\rm eff}(q^2)$.
We ignore non-factorizable effects, which are expected to play a non-negligible role, particularly
at large hadronic recoil \cite{Beneke:2001at,Beneke:2004dp}.

Notice that the radiative operators $\op{7(')}$ contribute to the semi-leptonic decay
through photon exchange and its electromagnetic coupling to a lepton pair
such that
\begin{align}
    \bra{\Lambda(k) \ell^+(q_1) \ell^-(q_2)} \op{7(')} \ket{\Lambda_b(p)}
    & = -\frac{2 m_b}{q^2} \, \bra{\Lambda} \bar{s} \, i\sigma^{\mu\nu} q_\nu \, P_{R(L)} b \ket{\Lambda_b} \,
    [\bar{u_\ell} \gamma_\mu v_\ell]\,,
\end{align}
with $q^\mu=p^\mu-k^\mu$ being the momentum transfer to the lepton pair.
In this notation the Wilson coefficient $\wilson{7'}$ is suppressed by $m_s/m_b$ in the SM.

Potential NP contributions to $b\to s\gamma$ and $b\to s\ell^+\ell^-$, on the one hand,
modify the Wilson coefficients of the SM operators above and, on the other hand,
feed into new effective operators,
\begin{equation}
    {\cal H}^\text{eff} \Big|_{\rm NP, naive}
    = N_1 \, \sum_{i=S('),P('),9',10',T,T5} \wilson{i}(\mu) \, \op{i}
    \,.
\end{equation}
Here the SM operators are completed to form a basis of dimension-six operators by:
\begin{align}
    \op{S(')}
    & = \[\bar s P_{R(L)} b\]\[\bar\ell \ell]\,, &
    \op{P(')}
    & = \[\bar s P_{R(L)} b\]\[\bar\ell \gamma_5\ell]\,,\cr
    \op{9'}
    & = \[\bar s \gamma^\mu P_{R} b\]\[\bar\ell \gamma_\mu \ell\]\,, &
    \op{10'}
    & = \[\bar s \gamma^\mu P_{R} b\]\[\bar\ell \gamma_\mu \gamma_5 \ell\]\,,\cr
    \op{T}
    & = \[\bar s \sigma^{\mu\nu} b\] \[\bar\ell \sigma_{\mu\nu}\ell\]\,, &
    \op{T5}
    & = \frac{i}{2}  \eps_{\mu\nu\alpha\beta} \,
    \[\bar s \sigma^{\mu\nu} b\] \[\bar\ell \sigma^{\alpha_\beta}\ell\]\,, &
\end{align}
which contains the chirality-flipped counterparts of the SM operators $\op9$ and
$\op{10}$ together with scalar, pseudoscalar, tensor and pseudotensor operators.

%%
%% Decay
%%
\section{The Decays $\Lambda_b \to \Lambda(\to N \pi) \ell^+\ell^-$}

\label{sec:decay}

\subsection{Kinematics}
\label{sec:decay:kin}

\begin{figure}
\centering
    \includegraphics[width=.7\textwidth]{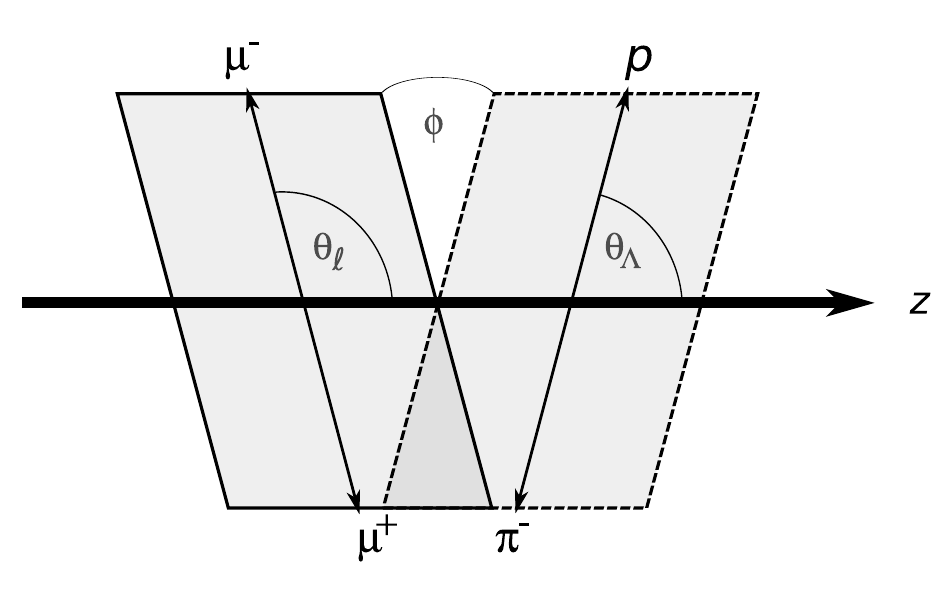}
    \caption{\label{fig:kin} Topology of the decay $\Lambda_{b} \to \Lambda\,(\to p \pi^-) \, \ell^+\ell^-$.}
\end{figure}

We assign particle momenta and spin variables for the baryonic states
in the decay according to:
\begin{eqnarray}
    \label{eq:hadkin}
    \Lambda_b(p, s_{\Lambda_b}) &\to& \Lambda(k, s_\Lambda) \, \ell^+(q_1) \, \ell^-(q_2)\,,
    \cr
    \Lambda(k, s_\Lambda) & \to &  N(k_1, s_N)\, \pi(k_2) \,, \qquad \left(N\pi = \{p\pi^-,n\pi^0\}\right)\,.
\end{eqnarray}
Here, $s_i$ are the projections of the baryonic spins onto the $z$-axis in their rest frames.
It is convenient to define sums and differences of the hadronic and leptonic momenta
in the final state,
\begin{align}
    q^\mu                   & = q_1^\mu  + q_2^\mu \,, &
    \bar{q}^\mu              & = q_1^\mu  - q_2^\mu \,, &
    k^\mu                   & = k_1^\mu  + k_2^\mu \,, &
    \bar{k}^\mu              & = k_1^\mu  - k_2^\mu \,.
\end{align}
The discussion of the kinematics is similar to what has been already worked out
for semileptonic four-body $B$-meson decays, see e.g.~\cite{Lee:1992ih,Faessler:2002ut,Kruger:2005ep,Altmannshofer:2008dz,Faller:2013dwa}.
We  end up with four independent kinematic variables, which can be chosen as
the invariant mass $q^2$, the helicity angles $\theta_\Lambda$ and $\theta_\ell$,
and the azimuthal angle $\phi$, see Fig.~\ref{fig:kin}, which are defined in the relevant Lorentz frames in \refapp{kin}.

Furthermore, we introduce a set of virtual polarization vectors
$\eps^\mu(\lambda= t,+,-,0)$ with $q\cdot \eps(\pm) = q\cdot \eps(0) = 0$, which in the dilepton rest frame
take the canonical form as shown in \refapp{kin}.
In terms of the leptonic angle $\theta_\ell$ and the relativistic lepton velocity,
$$
 \beta_\ell = \sqrt{1- \frac{4m_\ell^2}{q^2}} \,,
$$
the polarization vectors obey
\begin{equation}
\begin{aligned}
    \bar q \cdot \eps(t)                 & = 0 \,, &
    \bar q \cdot \eps(\pm)               & = \pm\frac{\beta_\ell}{\sqrt{2}} \,\sqrt{q^2} \,\sin\theta_\ell \,, &
    \bar q \cdot \eps(0)                 & = -\beta_\ell \, \sqrt{q^2} \, \cos\theta_\ell\,.
\end{aligned}
\end{equation}
Similarly, the corresponding hadronic kinemetic variables appear in
the following Lorentz-scalar products,
%The remaining angle can be extracted from the following Lorentz invariant quantities
\begin{equation}
\begin{aligned}
    \eps^*(t)   \cdot \bar k    & = \frac{\beta_{N\pi}}{2} \sqrt{\lambda} \cos\theta_\Lambda\,,\qquad
    \eps^*(\pm) \cdot \bar{k}   & = \pm \frac{\beta_{N\pi}}{\sqrt{2}} m_\Lambda \sin\theta_\Lambda \exp\big(\pm i \phi \big)\,.
\end{aligned}
\end{equation}
Here, we abbreviate $\lambda \equiv \lambda(\mlamB^2, \mlam^2, q^2)$, with
the K\"all\'en function
\begin{equation}
    \lambda(a, b, c) = a^2 + b^2 + c^2 - 2 (a b + a c + b c)\,.
\end{equation}
We also define
\begin{equation}
    \beta_{N\pi} = \frac{\sqrt{\lambda(\mlam^2, m_N^2, m_\pi^2)}}{\mlam^2}\,.
\end{equation}

\subsection{Hadronic Matrix Elements}
\label{sec:decay:hme}

\subsubsection{$\Lambda_b\to \Lambda$ Helicity Form Factors}

The hadronic form factors for $\Lambda_b \to \Lambda$ transitions are most conveniently
defined in the helicity basis \cite{Feldmann:2011xf}.
For the vector current, this yields three independent helicity form factors $f_i^V(q^2)$,
entering the corresponding helicity amplitudes $H_i^V(s_{\Lambda_b},s_\Lambda)$,
\begin{align}
H_t^V(s_{\Lambda_b}, s_\Lambda) & \equiv {\eps}_\mu^*(t) \,
\langle \Lambda(k, s_\Lambda) | \bar{s} \gamma^\mu  b | \Lambda_b(p, s_{\Lambda_b})\rangle
 \cr
 &= f_t^V(q^2) \, \frac{m_{\Lambda_b} - m_\Lambda}{\sqrt{q^2}}
 \left[ \bar u(k,s_\Lambda) \, u(p,s_{\Lambda_b}) \right]
    \label{eq:hel-ff-first}
 %h^S(s_{\Lambda_b}, s_\Lambda)
 \,,\cr
H_0^V(s_{\Lambda_b}, s_\Lambda) &\equiv {\eps}_\mu^*(0) \,
\langle \Lambda(k, s_\Lambda) | \bar{s} \gamma^\mu b | \Lambda_b(p, s_{\Lambda_b})\rangle
\cr & = 2 \, f^{V}_0(q^2) \, \frac{m_{\Lambda_b} + m_\Lambda}{s_+}
\, (k \cdot\eps^*(0)) \left[ \bar u(k,s_\Lambda) \, u(p,s_{\Lambda_b}) \right]
%[k\cdot \eps^*(0)] h^S(s_{\Lambda_b},s_\Lambda)
\,,\cr
H_\pm^V(s_{\Lambda_b}, s_\Lambda) &\equiv{\eps}_\mu^*(\pm) \,
\langle \Lambda(k, s_\Lambda) | \bar{s} \gamma^\mu b | \Lambda_b(p, s_{\Lambda_b})\rangle
 \cr &= f^{V}_\perp(q^2)
 \left[ \bar u(k,s_\Lambda) \, \slash \eps^*(\pm) \, u(p,s_{\Lambda_b}) \right]
 %[h^{V}(s_{\Lambda_b}, s_\Lambda) \cdot \eps^*(\pm)]
 \,,
\end{align}
where we slightly changed notation compared to \cite{Feldmann:2011xf}.
The kinematic functions $s_\pm$ are defined as
\begin{equation}
    s_\pm \equiv (m_{\Lambda_b} \pm m_\Lambda)^2 - q^2\,.
    \label{eq:spm}
\end{equation}
Analogous expressions are obtained for the axial-vector current,
\begin{align}
H_t^A(s_{\Lambda_b}, s_\Lambda) &\equiv {\eps}_\mu^*(t) \,
\langle \Lambda(k, s_\Lambda) | \bar{s}  \gamma^\mu\gamma_5 b | \Lambda_b(p, s_{\Lambda_b})\rangle
\cr
& = -f_t^A(q^2) \, \frac{m_{\Lambda_b} + m_\Lambda}{\sqrt{q^2}}
\left[ \bar u(k,s_\Lambda) \, \gamma_5 \, u(p,s_{\Lambda_b}) \right]
\,,\cr
H_0^A(s_{\Lambda_b}, s_\Lambda) &\equiv {\eps}_\mu^*(0)
\, \langle \Lambda(k, s_\Lambda) | \bar{s} \gamma^\mu \gamma_5 b | \Lambda_b(p, s_{\Lambda_b})\rangle
\cr & = -2 \, f^{A}_0(q^2) \, \frac{m_{\Lambda_b} - m_\Lambda}{s_-}
\, (k\cdot \eps^*(0))
\left[ \bar u(k,s_\Lambda) \, \gamma_5 \, u(p,s_{\Lambda_b}) \right]
\,,\cr
H_\pm^A(s_{\Lambda_b}, s_\Lambda) &\equiv{\eps}_\mu^*(\pm) \,
\langle \Lambda(k, s_\Lambda) | \bar{s} \gamma^\mu \gamma_5 b | \Lambda_b(p, s_{\Lambda_b})\rangle
\cr & = f^{A}_\perp(q^2)
\left[ \bar u(k,s_\Lambda) \, \slash \eps^*(\pm) \, \gamma_5 \, u(p,s_{\Lambda_b}) \right]
\,.
\end{align}
Restricting ourselves to the SM operator basis and its flipped counterpart,
only the $q^\nu$ projections of the tensor and pseudotensor currents appears,
which lead to another four independent form factors,
\begin{align}
H_0^T(s_{\Lambda_b}, s_\Lambda) &\equiv  \eps^*_\mu(0) \,
\langle \Lambda(k, s_\Lambda) | \bar{s} \, i \sigma^{\mu\nu} q_\nu \, b|\Lambda_b(p, s_{\Lambda_b})\rangle
\cr
& = -2 \, f_0^T(q^2) \, \frac{q^2}{s_+} \,
(k \cdot \eps^*(0))
\left[ \bar u(k,s_\Lambda) \, u(p,s_{\Lambda_b}) \right]
\,, \cr
H_\pm^T(s_{\Lambda_b}, s_\Lambda)& \equiv
\langle \Lambda(k, s_\Lambda) | \bar{s} \, i \sigma^{\mu\nu} q_\nu \,  b|\Lambda_b(p, s_{\Lambda_b})\rangle \eps^*_\mu(\pm)
\cr
& = - f_\perp^T(q^2) \, (\mlamB + \mlam)
\left[ \bar u(k,s_\Lambda) \,\slash \eps^*(\pm) \, u(p,s_{\Lambda_b}) \right]
\,,
\end{align}
and
\begin{align}
H_0^{T5}(s_{\Lambda_b}, s_\Lambda) &\equiv
\eps^*_\mu(0) \,
\langle \Lambda(k, s_\Lambda) | \bar{s} \, i \sigma^{\mu\nu} q_\nu \, \gamma_5 b|\Lambda_b(p, s_{\Lambda_b})\rangle
\cr
& = -2 \,f_0^{T5}(q^2) \, \frac{q^2}{s_-} \,(k \cdot \eps^*(0))
\left[ \bar u(k,s_\Lambda) \, \gamma_5 \, u(p,s_{\Lambda_b}) \right]
\,, \cr
H_\pm^{T5}(s_{\Lambda_b}, s_\Lambda) & \equiv
\eps^*_\mu(\pm) \,
\langle \Lambda(k, s_\Lambda) | \bar{s}  \, i \sigma^{\mu\nu} q_\nu  \, \gamma_5 b|\Lambda_b(p, s_{\Lambda_b})\rangle
\cr
& = f_\perp^{T5}(q^2) \, (\mlamB - \mlam)
\left[ \bar u(k,s_\Lambda) \, \slash \eps^*(\pm) \gamma_5 \, u(p,s_{\Lambda_b}) \right]
\,.
\label{eq:hel-ff-last}
\end{align}
The spinor matrix elements for given combinations of spin orientations are summarized
in \refapp{spinors}.
For the nonzero helicity amplitudes, we then obtain in
case of the vector current
\begin{align}
    H_t^V(+1/2, +1/2) = H_t^V(-1/2, -1/2)
        & = f_t^V(q^2) \, \frac{m_{\Lambda_b} - m_\Lambda}{\sqrt{q^2}} \, \sqrt{s_+}
        \,,\cr
    H_0^V(+1/2, +1/2) =  H_0^V(-1/2, -1/2)
        & = f_0^V(q^2)\, \frac{m_{\Lambda_b} + m_\Lambda}{\sqrt{q^2}} \, \sqrt{s_-}
        \,,\cr
    H_+^V(-1/2, +1/2) =  H_-^V(+1/2, -1/2)
        & = -f_\perp^V(q^2) \, \sqrt{2 s_-}\,.
\end{align}
Similarly, in case of the axial-vector current we have
\begin{align}
    H_t^A(+1/2,+ 1/2) = - H_t^A(-1/2,- 1/2)
        & = f_t^A(q^2) \, \frac{m_{\Lambda_b} + m_\Lambda}{\sqrt{q^2}} \, \sqrt{s_-}\,,\cr
    H_0^A(+1/2, +1/2) =  - H_0^A(-1/2, -1/2)
        & = f_0^A(q^2) \, \frac{m_{\Lambda_b} - m_\Lambda}{\sqrt{q^2}} \, \sqrt{s_+}\,,\cr
    H_+^A(- 1/2, + 1/2) =  -H_-^A(+1/2, -1/2)
        & = - f_\perp^A(q^2) \, \sqrt{2 s_+}\,.
\end{align}
For the tensor and pseudotensor currents,
the non-vanishing helicity amplitudes read
\begin{align}
    H_0^T(+ 1/2, + 1/2) =  H_0^T(-1/2, -1/2)
        & = -f_0^T(q^2) \, \sqrt{q^2} \, \sqrt{s_-}\,,\cr
    H_+^T(- 1/2, + 1/2) =  H_-^T(+1/2, -1/2)
        & =f_\perp^T(q^2) \, (\mlamB + \mlam) \,  \sqrt{2 s_-}\,,
\end{align}
and
\begin{align}
    H_0^{T5}(+ 1/2, + 1/2) = -  H_0^{T5}(- 1/2, - 1/2)
        & =  f_0^{T5}(q^2) \, \sqrt{q^2} \, \sqrt{s_+}\,,\cr
    H_+^{T5}(- 1/2, + 1/2) = - H_-^{T5}(+ 1/2, - 1/2)
        & = - f_\perp^{T5}(q^2) \, (\mlamB - \mlam) \,  \sqrt{2 s_+}\,.
\end{align}
From these relations the advantage of using form factors defined in the
helicity basis becomes evident.\\

We combine the contributions of the individual operators with the
corresponding Wilson coefficients, distinguish the contributions for
different lepton chiralities, and change to the transversity basis.
The primary decay $\Lambda_b \to \Lambda \ell^+\ell^-$ is then
described by 8 transversity amplitudes, which we denote as
\begin{equation}
\begin{aligned}
    A_{\perp_1}^{L(R)} & = +\sqrt{2} N \left(C_{9,10,+}^{L(R)} H_+^V(-1/2,+1/2)
    - \frac{2 m_b \big(\wilson{7} + \wilson{7'}\big)}{q^2} H_+^{T }(-1/2,+1/2)\right)\,,\cr
    A_{\para_1}^{L(R)} & = -\sqrt{2} N \left(C_{9,10,-}^{L(R)} H_+^A(-1/2,+1/2)
    + \frac{2 m_b \big(\wilson{7} - \wilson{7'}\big)}{q^2} H_+^{T5}(-1/2,+1/2)\right)\,,\cr
    A_{\perp_0}^{L(R)} & = +\sqrt{2} N \left(C_{9,10,+}^{L(R)} H_0^V(+1/2,+1/2)
    - \frac{2 m_b \big(\wilson{7} + \wilson{7'}\big)}{q^2} H_0^{T }(+1/2,+1/2)\right)\,,\cr
    A_{\para_0}^{L(R)} & = -\sqrt{2} N \left(C_{9,10,-}^{L(R)} H_0^A(+1/2,+1/2)
    + \frac{2 m_b \big(\wilson{7} - \wilson{7'}\big)}{q^2} H_0^{T5}(+1/2,+1/2)\right)\,.
\end{aligned}
\end{equation}
Here, we abbreviate the various combinations of Wilson coefficients $\wilson{9,10}$ as
\begin{align}
    C_{9,10,+}^{L(R)} & = (\wilson{9} \mp \wilson{10}) + (\wilson{9'} \mp \wilson{10'})\,, &
    C_{9,10,-}^{L(R)} & = (\wilson{9} \mp \wilson{10}) - (\wilson{9'} \mp \wilson{10'})\,.
\end{align}

The normalization factor $N$,
\begin{equation}
    N
    = N_1 \, \sqrt{\frac{q^2 \, \sqrt{\lambda(\mlamB^2, \mlam^2, q^2)}}{3\cdot 2^{10} \, \mlamB^3 \,\pi^3}}
    = \gfermi \, V_{tb} V_{ts}^* \, \alpha_e \,
    \sqrt{\frac{q^2 \, \sqrt{\lambda(\mlamB^2, \mlam^2, q^2)}}{3\cdot 2^{11} \, \mlamB^3 \, \pi^5}}\,,
\end{equation}
is chosen so that $\dd \Gamma = \sum_\lambda |A_\lambda|^2 \dd q^2$.
This generalizes the parametrization in terms of transversity amplitudes
(for instance as described in \cite{Bobeth:2012vn})
to exclusive fermionic $b\to s\ell^+\ell^-$ transitions.

\subsubsection{Hadronic Couplings in $\Lambda\to N\pi$}

In the SM the decay $\Lambda  \to N\pi$ is described by the effective Hamiltonian
\begin{equation}
    H^\text{eff}_{\Delta S = 1} =\underbrace{ \frac{4 \gfermi}{\sqrt{2}} \, V_{ud}^* V_{us}}_{N_2}
    \left[\bar{d} \gamma_\mu P_L u\right]\left[\bar{u} \gamma^\mu P_L s\right] \,.
\end{equation}
The hadronic matrix element which determines the $\Lambda \to N\pi$ decay
can be parametrized as \cite{Okun:1965}
\begin{align}
&
\bra{p(k_1, s_N) \pi^-(k_2)} \left[\bar{d} \gamma_\mu P_L u\right]\left[\bar{u} \gamma^\mu P_L s\right]
\ket{\Lambda(k, s_\Lambda)} \cr
& = \big[\bar u(k_1, s_N) \big(\xi \,\gamma_5 + \omega\big) u(k, s_\Lambda)\big] \equiv H_2(s_\Lambda, s_N)\,.
\end{align}
As a consequence of the equations of motion, only two independent hadronic parameters
appear which we have denoted as $\xi$ and $\omega$.\footnote{In the notation of Okun \cite{Okun:1965}
our convention translates as $\omega=\alpha^{\rm Okun}$ and $\xi=-\beta^{\rm Okun}$. The corresponding $\Lambda \to n\pi^0$
decay parameters are related by isospin symmetry, neglecting electromagnetic and light quark-mass
effects.}
They can be extracted from the $\Lambda \to p \pi^-$ decay width and polarization measurements.

In terms of the kinematic variables introduced above (see also \refapp{kin}), the helicity
amplitudes for the secondary decay can be written as
\begin{align}
    H_2(+1/2,+1/2) &= \left(\sqrt{r_+} \, \omega - \sqrt{r_-} \, \xi \right) \cos\frac{\theta_\Lambda}{2} \,, \cr
    H_2(+1/2,-1/2) &= \left(\sqrt{r_+} \, \omega + \sqrt{r_-} \, \xi \right) \sin\frac{\theta_\Lambda}{2}\,
    e^{i \phi}  \,, \cr
    H_2(-1/2,+1/2) &= \left(-\sqrt{r_+} \, \omega + \sqrt{r_-} \, \xi \right) \sin\frac{\theta_\Lambda}{2}\,
    e^{-i \phi} \,, \cr
    H_2(-1/2,-1/2) &= \left(\sqrt{r_+}\, \omega + \sqrt{r_-} \,\xi \right) \cos\frac{\theta_\Lambda}{2} \,.
\end{align}
where we abbreviate
\begin{equation}
    r_{\pm} \equiv (m_{\Lambda} \pm m_N)^2 - m_{\pi}^2\,.
\end{equation}
The corresponding helicity contributions to the decay width can be defined as
\begin{equation}
    \Gamma_2(s_{\Lambda}^{(a)},s_{\Lambda}^{(b)}) =  |N_2|^2 \frac{\sqrt{r_+ r_-}}{16 \pi m_{\Lambda}^3} \,
    \sum_{s_N} H_2(s_{\Lambda}^{(a)},s_N) \, H_2^* (s_{\Lambda}^{(b)},s_N) \,,
\end{equation}
which yield
\begin{align}
     \Gamma_2(+1/2, +1/2) &=  (1 + \alpha  \, \cos\theta_\Lambda) \, \Gamma_\Lambda\,,\qquad
     \Gamma_2(+1/2, -1/2) = -\alpha \, \sin\theta_\Lambda \, e^{i \phi} \, \Gamma_\Lambda \,,\cr
    \Gamma_2(-1/2, -1/2) &= (1 - \alpha \, \cos\theta_\Lambda) \, \Gamma_\Lambda \,,\qquad
    \Gamma_2(-1/2, +1/2) = -\alpha \, \sin\theta_\Lambda e^{-i \phi}\, \Gamma_\Lambda \,.
\end{align}
Here the $\Lambda \to N\pi$ decay width is given as \cite{Okun:1965}
\begin{align}
    \Gamma_\Lambda & =  \Gamma_2(+1/2, +1/2) + \Gamma_2(-1/2, -1/2)
                    = \frac{|N_2|^2 \sqrt{r_+ r_-}}{16 \pi m_\Lambda^3} \left(
                    r_- \,|\xi|^2 + r_+ \, |\omega|^2 \right)\,,
\end{align}
and the parity-violating decay parameter $\alpha$ reads
\begin{equation}
    \alpha = \frac{-2 \,\Re{\omega \,\xi}}{\sqrt{\frac{r_-}{r_+}} \, |\xi|^2 + \sqrt{\frac{r_+}{r_-}} \, |\omega|^2} = +\alpha^{\text{exp}}\,.
\end{equation}

\subsection{Angular Observables}
\label{sec:decay:obs}

The angular distribution for the 4-body decay can be written
as a 4-fold differential decay width,
\begin{equation}
  K(q^2, \cos\theta_\ell, \cos\theta_\Lambda, \phi) \equiv
  \frac{8\pi}{3} \frac{\dd^4 \Gamma}{\dd q^2\,\dd \cos\theta_\ell\,\dd \cos\theta_\Lambda\,\dd \phi} \,,
\end{equation}
which can be decomposed in terms of a set of trigonometric functions,
\begin{equation}
\begin{aligned}
    \label{eq:angular-distribution}
    K(q^2, \cos\theta_\ell, \cos\theta_\Lambda, \phi)
    & =
         \big( K_{1ss} \sin^2\theta_\ell +\, K_{1cc} \cos^2\theta_\ell + K_{1c} \cos\theta_\ell\big) \,\cr
    &  + \big( K_{2ss} \sin^2\theta_\ell +\, K_{2cc} \cos^2\theta_\ell + K_{2c} \cos\theta_\ell\big) \cos\theta_\Lambda
    \cr
    &  + \big( K_{3sc}\sin\theta_\ell \cos\theta_\ell + K_{3s} \sin\theta_\ell\big) \sin\theta_\Lambda \sin\phi\cr
    &  + \big( K_{4sc}\sin\theta_\ell \cos\theta_\ell + K_{4s} \sin\theta_\ell\big) \sin\theta_\Lambda \cos\phi \,.
\end{aligned}
\end{equation}
Here the first line corresponds to a relative angular momentum $(L,M)$ between the $N\pi$ system and the dilepton system of
$(L,M) = (0,0)$. The lines two to four correspond to $L = 1$, with the third component $M = 0$ in the second line,
and $|M|=1$ in lines three and four. This implies that each line of
\refeq{angular-distribution} can be decomposed in terms of associated Legendre polynomials $P_l^{|M|}(\cos\theta_\ell)$,
where $0 \leq l \leq 2$ holds for the dilepton angular momentum $l$ on the basis of angular momentum conservation.
This agrees exactly with our results \refeq{angular-distribution}. In particular,
\begin{itemize}
    \item[(a)] there are no terms $\propto \sin\theta_\ell(\cos\theta_\ell)$ or $\propto \sin\theta_\ell(\cos\theta_\ell)\cos\theta_\Lambda$,
    \item[(b)] there are no terms $\propto \sin^2\theta_\ell \sin\theta_\Lambda$, $\propto \cos\theta_\ell\sin\theta_\Lambda$ or $\propto \cos^2\theta_\ell\sin\theta_\Lambda$, and
    \item[(c)] no further terms can arise from dimension-six operators which are absent in our calculation; i.e., scalar and tensor operators.
\end{itemize}

The coefficients in the decomposition \refeq{angular-distribution} are refered to as
angular observables and depend on the dilepton invariant mass.
In our notation, they are denoted as
$K_{n\lambda} \equiv K_{n\lambda}(q^2)$, with $n=1,\dots,4$, and
$\lambda = s,c,ss,cc,sc$. In terms of the transversity
amplitudes for $\Lambda_b \to \Lambda$ transitions and the decay parameter $\alpha$ in $\Lambda \to N\pi$
defined above, we find
\begin{align}
% 1
    K_{1ss}(q^2) & = \frac{1}{4}\Big[|A_{\perp_1}^R|^2 + |A_{\para_1}^R|^2
    + 2 |A_{\perp_0}^R|^2 + 2 |A_{\para_0}^R|^2 + (R \leftrightarrow L)\Big] \,, \cr
    K_{1cc}(q^2) & = \frac{1}{2}\Big[|A_{\perp_1}^R|^2 + |A_{\para_1}^R|^2 + (R \leftrightarrow L)\Big]\,, \cr
    K_{1c}(q^2)  & = -\Re{A_{\perp_1}^R A_{\para_1}^{*R} - (R \leftrightarrow L)}
\end{align}
and
\begin{align}
% 2
    K_{2ss}(q^2) & = +\frac{\alpha}{2}\Re{A_{\perp_1}^R A_{\para_1}^{*R}
    + 2 A_{\perp_0}^R A_{\para_0}^{*R} + (R \leftrightarrow L)} \,, \cr
    K_{2cc}(q^2) & = +\alpha\Re{A_{\perp_1}^R A_{\para_1}^{*R} + (R \leftrightarrow L)} \,, \cr
    K_{2c}(q^2)  & = -\frac{\alpha}{2}\Big[|A_{\perp_1}^R|^2 + |A_{\para_1}^R|^2 - (R \leftrightarrow L)\Big] \,,
\end{align}
% 3
and
\begin{align}
    K_{3sc}(q^2) & = +\frac{\alpha}{\sqrt{2}}\Im{A_{\perp_1}^R A_{\perp_0}^{*R}
    - A_{\para_1}^{R} A_{\para_0}^{*R} + (R \leftrightarrow L)}\,, \cr
    K_{3s}(q^2)  & = +\frac{\alpha}{\sqrt{2}}\Im{A_{\perp_1}^R A_{\para_0}^{*R}
    - A_{\para_1}^{R} A_{\perp_0}^{*R} - (R \leftrightarrow L)}\,,
\end{align}
and
\begin{align}
% 4
    K_{4sc}(q^2) & = +\frac{\alpha}{\sqrt{2}}\Re{A_{\perp_1}^R A_{\para_0}^{*R}
    - A_{\para_1}^{R} A_{\perp_0}^{*R} + (R \leftrightarrow L)}\,, \cr
    K_{4s}(q^2)  & = +\frac{\alpha}{\sqrt{2}}\Re{A_{\perp_1}^R A_{\perp_0}^{*R}
    - A_{\para_1}^{R} A_{\para_0}^{*R} - (R \leftrightarrow L)} \,.
\end{align}
The angular observables contain all the relevant information about the
short- and long-distance dynamics in the SM or its extension by chirality-flipped
operators in the effective Hamiltonian.

\section{Phenomenological Applications}
\label{sec:pheno}

\subsection{Simple Observables}

For the experimental analyses
one can construct weighted angular integrals of the differential decay width,
\begin{equation}
    X(q^2) \equiv \int \frac{\dd^4 \Gamma}{\dd q^2\, \dd\cos\theta_\ell\, \dd\cos\theta_\Lambda\, \dd \phi}
    \, \omega_X(q^2,\cos\theta_\ell,\cos\theta_\Lambda,\phi) \, \dd \cos\theta_\ell \, \dd \cos\theta_\Lambda \, \dd \phi \,,
\end{equation}
to obtain different decay distributions in the dilepton invariant mass $q^2$
as linear combinations of angular observables (or ratios thereof).
\begin{itemize}
 \item[(a)] The simplest of these distributions is just the differential decay width in $q^2$,
 \begin{equation}
    \frac{\dd \Gamma}{\dd q^2}  = 2 K_{1ss} + K_{1cc} \,,
\end{equation}
which corresponds to $\omega_X\equiv 1$.

\item[(b)]
The fraction of transverse or longitudinal polarization of the dilepton system is obtained as
\begin{equation}
 F_1 = \frac{2K_{1cc}}{2 K_{1ss} + K_{1cc}} \,, \qquad
 F_0  = 1 - F_1  = \frac{2 K_{1ss} - K_{1cc}}{2 K_{1ss} + K_{1cc}} \,.
\end{equation}
This is achieved by the weight functions
\begin{equation}
\omega_{F_1} = \frac{5 \cos^2\theta_\ell - 1}{d\Gamma/dq^2} \,,
\qquad
\omega_{F_0} = \frac{2-5 \cos^2\theta_\ell}{d\Gamma/dq^2} \,.
\end{equation}

\item[(c)] The well-known forward-backward asymmetry with respect to the leptonic scattering angle,
normalized to the differential rate, is defined as
\begin{align}
    A^\ell_\text{FB} & =
    \frac{3}{2} \, \frac{K_{1c}}{2 K_{1ss} + K_{1cc}}    \,,
    &    \omega_{A^\ell_\text{FB}}          & = \frac{\sign [\cos\theta_\ell]}{d\Gamma/dq^2}
    \,.
\end{align}
\item[(d)] The analogous asymmetry for the bayonic scattering angle
reads
\begin{align}
  A^\Lambda_\text{FB}
  & = \frac{1}{2} \, \frac{2 K_{2ss} + K_{2cc}}{2 K_{1ss} + K_{1cc}}  \,,  &
  \omega_{A^\Lambda_\text{FB}}
  & = \frac{\sign [\cos\theta_\Lambda]}{d\Gamma/dq^2} \,.
\end{align}
\item[(e)]
Finally, one can also study a combined forward-backward asymmetry
via
\begin{align}
    A^{\ell\Lambda}_\text{FB} & = \frac{3}{4} \, \frac{K_{2c}}{2 K_{1ss} + K_{1cc}}
    &    \omega_{A^{\ell\Lambda}_\text{FB}} & = \frac{\sign [\cos\theta_\ell\,\cos\theta_\Lambda]}{d\Gamma/dq^2}\,.
\end{align}

\end{itemize}

\subsection{Exploiting Form-Factor Symmetries at Low Recoil}

In the limit of low hadronic recoil to the $(N\pi)$ system (i.e.\
large invariant lepton mass $q^2=\order{m_b^2} \gg \Lambda_{\rm QCD}^2$), the number of independent
form factors reduces as a consequence of the HQET spin symmetry
\cite{Mannel:1997xy}. In our notation the form-factor relation with the two leading
Isgur-Wise functions $\xi_1$ and $\xi_2$ read
(cf.\ \cite{Feldmann:2011xf})
\begin{align}
\xi_1-\xi_2 &=  f_\perp^V = f_0^V = f_\perp^T = f_0^T \,,
 \cr
\xi_1+\xi_2 &=  f_\perp^A = f_0^A = f_\perp^{T5} = f_0^{T5} \,.
 \label{eq:HQETrel}
\end{align}
In phenomenological analyses, one can make use of these relations in the following way:
\begin{itemize}
 \item At the moment, lattice-QCD estimates for the $\Lambda_b \to \Lambda$ form factors
   exist in the HQET limit \cite{Detmold:2012vy}. These results, together with their respective
   uncertainties, can be implemented as a theoretical prior distribution
   for a Bayesian analysis of the experimental data (within the OPE/factorization approximation).

  \item Measuring suitably defined combinations of angular observables (see below), one can
    obtain experimental information about the size of corrections to the form-factor relations
    \eqref{eq:HQETrel},
    in a similar way as has been proposed for the well-studied $B \to K^*(K\pi)\mu^+\mu^-$ decay
    \cite{Bobeth:2010wg,Bobeth:2012vn}.
    This will result in \emph{independent} posterior distributions for the individual
    form factors, as found in \cite{Hambrock:2012dg,Beaujean:2012uj,Hambrock:2013zya,Beaujean:2013soa}.

   \item For the time being,  priors for the individual form factors
     can be generated by estimates of Gaussian distributions
     for the subleading form factor contributions, which are defined in \refapp{hqet-beyond-leading-power}.
     In the long run,
     we expect lattice data for all individual form factors and their ratios.
\end{itemize}
In the following discussion, we will pursue a simplified approach that keeps the four
contributing vector and axialvector form factors as independent hadronic quantities,
and relates the four contributing tensor form factors via \eqref{eq:HQETrel},
including perturbative corrections at 1-loop (so called ``improved Isgur-Wise relations,
see \refapp{hqet-beyond-leading-power}).
In this approximation each transversity amplitude is proportional to
a single form factor and can be written as
\begin{align}
        A_{\perp_1}^{L(R)} & \simeq -2 \, N \, C_+^{L(R)} \, \sqrt{s_-}\ f_\perp^V(q^2) \,,
        \qquad
        A_{\para_1}^{L(R)}  \simeq  +2 \, N \, C_-^{L(R)}  \, \sqrt{s_+}\ f_\perp^A(q^2)\,,
\end{align}
and
\begin{align}
        A_{\perp_0}^{L(R)} & \simeq +\sqrt{2} \, N \, C_+^{L(R)} \, \frac{\mlamB + \mlam}{\sqrt{q^2}} \, \sqrt{s_-} \ f_0^V(q^2)\,,
        \cr
        A_{\para_0}^{L(R)} & \simeq -\sqrt{2} \,N \,C_-^{L(R)} \, \frac{\mlamB - \mlam}{\sqrt{q^2}} \, \sqrt{s_+} \ f_0^A(q^2)\,,
\end{align}
where we neglect $1/m_b$ corrections. The combinations of Wilson coefficients
that appear in this way are given by
\begin{align}
C_+^{R(L)} & = \Big((\wilson{9} + \wilson{9'}) + \frac{2 \kappa m_b \, \mlamB}{q^2}
(\wilson{7} + \wilson{7'}) \pm (\wilson{10} + \wilson{10'})\Big)\,,\\
C_-^{R(L)} & = \Big((\wilson{9} - \wilson{9'}) + \frac{2 \kappa m_b \, \mlamB}{q^2}
(\wilson{7} - \wilson{7'}) \pm (\wilson{10} - \wilson{10'})\Big)\,.
\end{align}
Here the parameter $\kappa=\kappa(\mu)$ contains the radiative QCD corrections to the
form factor relations such that together with the product of Wilson coefficients
and the $b$-quark mass the above expressions for the transversity amplitudes are
renormalization-scale independent (in a given order of perturbation theory).

The simplification for the transversity amplitudes directly translates to
the angular observables.
It turns out that these are sensitive to the following
combinations of short-distance parameters,
\begin{align}
  \rho_1^\pm &= \frac12 \left( |\wilson[R]{\pm}|^2 + |\wilson[L]{\pm}|^2 \right)
  = |\wilson{79}\pm\wilson{7'9'}|^2 + |\wilson{10}\pm\wilson{10'}|^2 \,, \cr
  \rho_2 &= \frac14 \left( \wilson[R]{+} \wilson[R*]{-} - \wilson[L]{-} \wilson[L*]{+} \right)
  \cr & = \Re{\wilson{79}\wilson[*]{10} - \wilson{7'9'}\wilson[*]{10'}}
  - i\Im{\wilson{79}\wilson[*]{7'9'} + \wilson{10}\wilson[*]{10'}} \,,
\end{align}
that also appear in the angular observables for  $B \to K^*(\to K\pi)\ell^+\ell^-$ decays \cite{Bobeth:2010wg,Bobeth:2012vn},
and two new bilinears of Wilson coefficients
\begin{align}
  \rho_3^\pm &= \frac12 \left( |\wilson[R]{\pm}|^2
  - |\wilson[L]{\pm}|^2 \right)= 2 \, \Re{(\wilson{79}\pm\wilson{7'9'}) (\wilson{10}\pm\wilson{10'})^*} \\
  \rho_4 &= \frac14 \left( \wilson[R]{+} \wilson[R*]{-} + \wilson[L]{-} \wilson[L*]{+} \right)
  \cr  & = \left( |\wilson{79}|^2 - |\wilson{7'9'}|^2 + |\wilson{10}|^2 - |\wilson{10'}|^2 \right)
  - i\Im{\wilson{79} \, \wilson[*]{10'} - \wilson{7'9'}\, \wilson[*]{10}} \,,
\end{align}
which contribute as a consequence of parity violation in the secondary weak decay. \footnote{%
    The combinations $\rho_3^-$ and $\rho_4$ also emerge for the non-resonant
    $B\to K\pi\ell^+\ell^-$ decays as recently found in \cite{Das:2014sra}.
}
Here we abbreviate
\begin{equation}
\begin{aligned}
    \wilson{79}   & \equiv \wilson[\text{eff}]{9} + \frac{2\kappa m_b \mlamB}{q^2} \wilson[\text{eff}]{7}\,, &
    \wilson{7'9'} & \equiv \wilson{9'} + \frac{2\kappa m_b \mlamB}{q^2} \wilson{7'}\,.
\end{aligned}
\end{equation}
We observe that in this approximation $K_{3sc} = 0$ even in the presence of chirally-flipped operators.

For the simple observables introduced in the previous subsection we then obtain
\begin{eqnarray}
\frac{\dd \Gamma}{\dd q^2}
 = 4 \, |N|^2  &&\left\lbrace \rho_1^+ \, s_- \left[2\, |f_\perp^V|^2
+ \frac{(\mlamB + \mlam)^2}{q^2} \, |f_0^V|^2\right] + \right.\cr &&  \left.  \rho_1^- \,
s_+ \left[2\, |f_\perp^A|^2 + \frac{(\mlamB - \mlam)^2}{q^2} \, |f_0^A|^2\right]\right\rbrace\,,
\end{eqnarray}
and
\begin{eqnarray}
   F_0
   =  4 \, |N|^2 && \left\lbrace\rho_1^+ \, s_- \frac{(\mlamB + \mlam)^2}{q^2} \, |f_0^V|^2 \right. \cr
 &&  \left. + \rho_1^- \, s_+ \frac{(\mlamB - \mlam)^2}{q^2} \, |f_0^A|^2\right\rbrace
  \left(\frac{\dd \Gamma}{\dd q^2} \right)^{-1}\,,
\end{eqnarray}
and
\begin{align}
     \frac{\dd \Gamma}{\dd q^2} \, A^\ell_\text{FB}          & =
     24 \, |N|^2 \, \Re{\rho_2} \, \sqrt{s_+ s_-} \, f_\perp^V f_\perp^A\,,\cr
    \frac{\dd \Gamma}{\dd q^2} \, A^\Lambda_\text{FB}       & = -8 \, |N|^2 \, \alpha \,
    \Re{\rho_4} \, \sqrt{s_+ s_-} \left\lbrace 2\,f_\perp^V f_\perp^A + \frac{\mlamB^2 - \mlam^2}{q^2} \,
    f_0^V f_0^A\right\rbrace\,,\cr
    \frac{\dd \Gamma}{\dd q^2} \, A^{\ell\Lambda}_\text{FB} & = -3 \,|N|^2 \,
    \alpha \left\lbrace\rho_3^+ \, s_- \, |f_\perp^V|^2 + \rho_3^- \, s_+ \, |f_\perp^A|^2\right\rbrace\,.
\end{align}
We will present numerical estimates for these observables in the SM
(integrated over $q^2$ in the low-recoil region)
in \refsec{numerics}.

Future experimental data will also allow to simultaneously test the short-distance
structure of the SM against NP, and to extract information on form-factor ratios.
In the presence of both SM-like and chirality-flipped operators,
we find one
ratio of angular observables where the form factors cancel
\emph{in the given approximation},
\begin{equation}
  \label{eq:def-x1}
  X_1 \equiv \frac{K_{1c}}{K_{2cc}} = -\frac{\Re{\rho_2}}{\alpha \, \Re{\rho_4}}\,,
\end{equation}
and two ratios of angular observables which only depend on form factors,
\begin{align}
  \frac{2 \, K_{2ss}}{K_{2cc}} & = 1 + \frac{\mlamB^2 - \mlam^2}{q^2} \, \frac{f_0^V f_0^A}{f_\perp^V f_\perp^A}\,,\cr
  \frac{2\, K_{4sc}}{K_{2cc}} & =
  \frac{\mlamB + \mlam}{\sqrt{q^2}} \, \frac{f_0^V}{f_\perp^V} - \frac{\mlamB - \mlam}{\sqrt{q^2}} \,
  \frac{f_0^A}{f_\perp^A}\,.
\end{align}
We also find ratios that are only functions of the Wilson coefficients
and a single ratio of form factors, $f_\perp^V/f_\perp^A$,
\begin{align}
  \frac{4\,K_{1cc}}{K_{1c}} & = \sqrt{\frac{s_-}{s_+}} \, \frac{\rho_1^+}{\Re{\rho_2}} \, \frac{f_\perp^V}{f_\perp^A}
  + \sqrt{\frac{s_+}{s_-}} \, \frac{\rho_1^-}{\Re{\rho_2}} \, \frac{f_\perp^A}{f_\perp^V}\,,\cr
  \frac{4\, K_{2c}}{K_{2cc}} & = \sqrt{\frac{s_-}{s_+}} \, \frac{\rho_3^+}{\Re{\rho_4}} \,\frac{f_\perp^V}{f_\perp^A}
  + \sqrt{\frac{s_+}{s_-}} \, \frac{\rho_3^-}{\Re{\rho_4}} \, \frac{f_\perp^A}{f_\perp^V}\,.
\end{align}

If, on the other hand, we assume the absence of chirality-flipped operators
(including $\wilson{7'} \to 0$ in the SM), we obtain
\begin{equation}
  \rho_3^\pm \to 2 \, \rho_2 = 2 \,
  \Re{\wilson{79}\wilson[*]{10}} \quad\text{and}\quad
  \rho_4 \to \frac12 \, \rho_1 = \frac12 \left(|\wilson{79}|^2 + |\wilson{10}|^2\right) .
\end{equation}
This further implies $K_{3s} \to 0$,
and we also finds one more ratio of observables which is free of form factors,
\begin{align}
  \label{eq:def-x1-x2}
  X_1                               & \to -\frac{2 \, \rho_2}{\alpha \, \rho_1}\,, &
  X_2 \equiv \frac{K_{1cc}}{K_{2c}} & = -\frac{2 \, \alpha \, \rho_2}{\rho_1}\,.
\end{align}
The additional short-distance free observables in the SM operator basis are
given by
\begin{align}
   \frac{2 \, K_{1ss} - K_{1cc}}{K_{1cc}} & =
   \frac{s_- \, (\mlamB+\mlam)^2 \, |f_0^V|^2 +
   s_+ \, (\mlamB-\mlam)^2 \, |f_0^A|^2}{q^2 \, s_- \, |f_\perp^V|^2 + q^2 \, s_+ \, |f_\perp^A|^2} \,,
   \cr
   \frac{2 \, K_{1ss} - K_{1cc}}{2K_{2ss} - K_{2cc}} &
   = -\frac{1}{2\alpha} \left( \frac{f_0^V}{f_0^A}  \, \sqrt{\frac{s_-}{s_+}} \,
     \frac{\mlamB+\mlam}{\mlamB-\mlam}
   + \frac{f_0^A}{f_0^V} \, \sqrt{\frac{s_+}{s_-}} \, \frac{\mlamB-\mlam}{\mlamB+\mlam} \right) \,,\cr
   \alpha^2 \, \frac{K_{1cc}}{K_{2cc}} = \frac{K_{2c}}{K_{1c}} & =
   -\frac{\alpha}{2} \left( \sqrt{\frac{s_-}{s_+}} \,
   \frac{f_\perp^V}{f_\perp^A} + \sqrt{\frac{s_+}{s_-}} \, \frac{f_\perp^A}{f_\perp^V} \right) \,.
\end{align}

\subsection{Simplifications at Large Recoil}

The number of independent form factors  reduces further in the
limit of large recoil energy \cite{Charles:1998dr,Beneke:2000wa} where the leading contributions\footnote{More precisely,
one has to distinguish soft overlap contributions  \cite{Mannel:2011xg,Feldmann:2011xf}
and hard spectator interactions \cite{Wang:2011uv}. In contrast to the analogous
$B$-meson transitions, the former are suppressed by one power of the $b$-quark
mass compared to the latter. On the other hand, the hard spectator term now only
starts at second order in the strong coupling $\alpha_s$ and therefore numerically
appears to be a sub-dominant effect. In any case, both contributions fulfill the
form-factor symmetry relations to first approximation.
}
can be identified using soft-collinear
effective theory (SCET \cite{Bauer:2000yr,Beneke:2002ph}). In our notation this simply implies
the equality of all helicity form factors,
\begin{align}
  f_\perp^V = f_0^V = f_\perp^T = f_0^T =  f_\perp^A = f_0^A = f_\perp^{T5} = f_0^{T5} \,.
 \label{eq:SCETrel}
\end{align}
Including $\alpha_s$ corrections to the soft form factors, the
modification of the form-factor relations \eqref{eq:SCETrel} can be described
by a vertex factor, and a single $q^2$-dependent function $\Delta \xi_\Lambda$ that emerges
from spectator scattering, see
appendix~C of \cite{Feldmann:2011xf}.
The values of $\xi_\Lambda$ and $\Delta \xi_\Lambda$ have been estimated
from sum rules with $\Lambda_b$ distribution amplitudes in \cite{Feldmann:2011xf}.
Due to the complexity of the baryonic transition and -- compared to the mesonic case --
the poor theoretical knowledge on the baryonic wave functions, these estimates have
a large uncertainty. With sufficient experimental information, however, one can
again try to constrain the form-factor values from the data itself.

To this end, we first recall that
the function $\Delta \xi_\Lambda$ drops out in the following sums,
\begin{align}
\frac{f_0^V + f_0^A}{2} \equiv \xi_\Lambda
\quad
\Rightarrow
\quad
\frac{f_\perp^V + f_\perp^A}{2}  &\simeq \left( 1 + \frac{\alpha_s C_F}{4\pi} \, L \right) \xi_\Lambda \,,
\cr
\frac{f_0^{T} + f_0^{T5}}{2} &\simeq  \left( 1 + \frac{\alpha_s C_F}{4\pi} \left( \ln \frac{m_b^2}{\mu^2} - 2 + 2L) \right) \right)
\xi_\Lambda \,,
\cr
\frac{f_\perp^{T} + f_\perp^{T5}}{2} &\simeq \left( 1 + \frac{\alpha_s C_F}{4\pi} \left( \ln \frac{m_b^2}{\mu^2} - 2 \right)  \right)
\xi_\Lambda \,.
\label{eq:SCETmod}
\end{align}
where $L = \frac{q^2-m_b^2}{q^2} \, \ln \left(1-\frac{q^2}{m_b^2}\right)$, and
the $\mu$-dependence of the tensor form factors related to the anomalous dimension
of the tensor operators is explicit. The corresponding \emph{differences}
between form factors will be proportional to $\Delta \xi_\Lambda$.
We remind the reader that in the
large-recoil region we also expect sizeable corrections from non-factorizable
(i.e.\ not form-factor like) contributions, which formally enter at the same order as
$\Delta \xi_\Lambda$. These have not been calculated or estimated at present, but
would be required for a consistent extraction of $\Delta \xi_\Lambda$.
The phenomenological strategy would then be the following.
\begin{itemize}
 \item Take \eqref{eq:SCETmod} as a theoretical constraint on the form factors.
 \item Use $f_0^V-f_0^A  \to f_\perp^V-f_\perp^A \to f_0^T-f_0^{T5} \to f_\perp^T - f_\perp^{T5} \to 0$
    as the central value for a theory prior on the form-factor differences with a conservative
    estimate for the theory uncertainty (say, of the order 20-30\%, independently for each
    individual form factor difference).
\item Based on comparison with upcoming experimental data, posterior predictive distributions
    for the form-factor differences $f_0^V - f_0^A$ and $f_\perp^V - f_\perp^A$ can be obtained.
    (Further differences could be constrained if data permits).
    These would indicate the size of corrections to the large-recoil symmetry relations. Using an
    explicit ansatz for both the factorizable and non-factorizable spectator effects, the size of
    these effects could be estimated.
\end{itemize}

As a first step, we may ignore the spectator effects altogether.
Including the known factorizing contributions from the relevant
hadronic operators in
$b\to s\ell^+\ell^-$ processes \cite{Asatryan:2001zw},
the transversity amplitudes can then
be written as
\begin{align}
        A_{\perp_1}^{L(R)} & = -2 N \left[ C_{9,10,+}^{L(R)} + \frac{2 m_b \, \mlamB}{q^2} \, \tau_{1,+}\right] f_\perp^V \, \sqrt{s_-}\,,\cr
        A_{\para_1}^{L(R)} & = +2 N \left[ C_{9,10,-}^{L(R)} + \frac{2 m_b \, \mlamB}{q^2} \, \tau_{1,-}\right] f_\perp^A \, \sqrt{s_+}\,,\cr
        A_{\perp_0}^{L(R)} & = +\sqrt{2} N \left[ C_{9,10,+}^{L(R)} + \frac{2 m_b}{\mlamB} \, \tau_{0,+}\right] f_0^V \, \sqrt{s_-}\,,\cr
        A_{\para_0}^{L(R)} & = -\sqrt{2} N \left[ C_{9,10,-}^{L(R)} + \frac{2 m_b}{\mlamB} \, \tau_{0,-}\right] f_0^A \, \sqrt{s_+}\,,
\end{align}
where the quantities $\tau_{i,\pm}$ can be expressed in terms of Wilson coefficients $\wilson{i}$, form-factor ratios $R_i$ and
perturbative functions $F_i^{(7,9)}$,
    \begin{align}
        \tau_{1,\pm} & = \frac{\mlamB \pm \mlam}{\mlamB} \left(\wilson[\rm eff]{7} \pm \wilson{7'} - \frac{\alpha_s}{4\pi} \sum_{i=1,2,8} \wilson{i} F_i^{(7)}(q^2, m_b, m_c)\right) R_1
                         \cr & \quad + \frac{q^2}{2 m_b \mlamB} \left(Y_9(q^2) - \frac{\alpha_s}{4\pi} \sum_{i=1,2,8} \wilson{i} F_i^{(9)}(q^2, m_b, m_c)\right)\,,\cr
        \tau_{0,\pm} & = \frac{\mlamB}{\mlamB \pm \mlam} \left(\wilson[\rm eff]{7} \pm \wilson{7'} - \frac{\alpha_s}{4\pi} \sum_{i=1,2,8} \wilson{i} F_i^{(7)}(q^2, m_b, m_c)\right) R_0
                        \cr & \quad + \frac{\mlamB}{2 m_b} \left(Y_9(q^2) - \frac{\alpha_s}{4\pi} \sum_{i=1,2,8} \wilson{i} F_i^{(9)}(q^2, m_b, m_c)\right)\,.
   \label{eq:correlator-large-recoil}
   \end{align}
and the function $Y_9(q^2)$ captures the one-loop virtual quark-loop contribution
which can be absorbed into $\wilson{9} \to \wilson[\rm eff]{9}(q^2)$.
The form-factor ratios occuring in this limit are simply given by
\begin{align}
        R_1 = \frac{f_\perp^{T}}{f_\perp^{V}}
        = \frac{f_\perp^{T5}}{f_\perp^{A}}
        & = 1 + \frac{\alpha_s C_F}{4\pi}\left(\ln\left(\frac{m_b^2}{\mu^2} - 2 - L\right)\right)\,,
 \cr
        R_0 = \frac{f_0^{T}}{f_0^{V}}   =
        \frac{f_0^{T5}}{f_0^{A}}         &
        = 1 + \frac{\alpha_s C_F}{4\pi}\left(\ln\left(\frac{m_b^2}{\mu^2} - 2 + 2L\right)\right)\,,
\end{align}

The three observable forward-backward asymmetries, $A_{\rm FB}^{\ell,\Lambda,\ell\Lambda}(q^2)$,
develop a characteristic $q^2$-behaviour (see numerical discussion below). 
In particular, we find that within the SM $A_{\rm FB}^{\ell}(q^2)$ and $A_{\rm FB}^{\ell\Lambda}(q^2)$
cross zero, where to first approximations the roots $q^2_{0,\ell}$ and $q^2_{0,\ell\Lambda}$
are the same, 
\begin{equation}
    q^2_{0,\ell} \simeq q^2_{0,\ell\Lambda} \simeq - 2 \, m_b \, \mlamB \, \frac{\wilson{7}}{\wilson{9}}\,.
\end{equation}
This expression is well known from other exclusive and the inclusive $b\to s\ell^+\ell^-$ decays.
On the other hand, $A_{\rm FB}^\Lambda(q^2)$ does not cross zero in the SM.

\subsection{Numerical Analysis}
\label{sec:numerics}

In order to translate our theoretical results into numerical predictions, 
we are going to determine predictive probability distributions for the
$\Lambda_b\to \Lambda(\to N\pi)\ell^+\ell^-$  observables. 
All central values and uncertainty ranges that we quote are
based on these distributions. In the course of our work, we extend
the EOS flavor program \cite{EOS} through implementation of the relevant
decay observables at both large and low hadronic recoil.

\subsubsection{Transition Form Factors}

Within our simplified factorization approach, the essential hadronic input functions
are the $\Lambda_b\to \Lambda$ transition form factors. For the numerical analysis 
we need probability distributions for the individual helicity form-factor values at different
values of $q^2$. To this end we take into account the available lattice data 
in the heavy-quark limit and an estimate of the soft form factor from a sum rule 
in the large-recoil limit, which are combined using the parametrization \refeq{ff-param}.
Corrections to the heavy-quark limit are allowed for as well.

The specific steps of our analysis aim toward a Bayesian analysis once
the knowledge of the transition form factors improves. These steps are presented
in detail in \refapp{ff-fit}.

Our setup includes eight form factor parameters $\vec{x}$, see \refeq{ff-fit:params}.
The main result of the fits is the \emph{posterior} $P(\vec{x} | \text{Estimates})$, which is defined
in \refeq{ff-fit:posterior}. The posterior is central to the computation of the numerical
results that follow.

\subsubsection{Results}

Theory uncertainties in the computation of the observables arise, beyond the transition form factors,
also from variations of CKM matrix elements, the $\Lambda\to N\pi$ coupling $\alpha$,
and the masses of the charm and bottom quark; see \reftab{num-input} for a summary.\\

For the CKM matrix elements, we use uncorrelated Gaussian distributions for our analysis.
Their parameters follow from the marginalised posterior distributions as obtained by the
UTfit collaboration in their ``Tree Level Fit'' analysis \cite{Bona:2006ah}. For $\alpha$
and the quark masses, we use world averages as provided by the Particle Data Group
\cite{Beringer:1900zz}. Further input parameters, such as the hadron masses and lifetimes, are fixed to the
central values of their respective world averages \cite{Beringer:1900zz}.
For the $\Lambda_b$ lifetime, we use the more recent world average by the Heavy Flavor
Averaging Group \cite{Amhis:2012bh},
\begin{equation}
    \tau_{\Lambda_b}^\text{HFAG} = 1.451 \pm 0.013\,\text{ps}\,,
\end{equation}
which includes the two recent measurements by the LHCb collaboration \cite{Aaij:2013oha,Aaij:2014owa}.\\

The observables of interest in our analysis are the branching ratio $\mathcal{B}$, the three forward-backward
asymmetries $A_{\rm FB}^\ell$, $A_{\rm FB}^\Lambda$ and $A_{\rm FB}^{\ell\Lambda}$, and the fraction of longitudinal lepton pairs $F_0$.
For the numerical evaluation, we first calculate $q^2$-integrated angular observables $K_{n\lambda}$. All observables of interest are then computed
from these pre-integrated angular observables, which we denote as $\langle O\rangle$ for any observable $O(q^2)$.
Our nominal choice of the integration region is $15\,\GeV^2 \leq q^2 \leq (\mlamB - \mlam)^2$,
in order to minimize the uncertainties from quark-hadron duality violation
(see the discussion for the mesonic counterpart in \cite{Beylich:2011aq}).\footnote{%
    Note that in view of the present experimental situation \cite{Aaij:2013pta} the effects from
    quark-hadron duality violation can still be sizable, even after averaging over the entire
    low-recoil region. A local description of the spectrum, similar to the mesonic case as discussed in \cite{Khodjamirian:2010vf,Khodjamirian:2012rm,Lyon:2014hpa},
    is beyond the scope of this work.
}

We obtain numerical estimates for the observables $\vec{O}$ through uncertainty propagation. For this,
we compute $8\cdot 10^5$ variates of the predictive distribution $P(\vec{O})$,
\begin{equation}
    P(\vec{O}) = \iint \dd{\vec x}\,\dd{\vec\nu}\,\delta(\vec{O} - \vec{O}(\vec x,\vec\nu))\,P(\vec x|\text{Estimates})\,P_0(\vec\nu)\,.
\end{equation}
Here, $P_0(\vec\nu)$ is the prior of our nuisance parameters $\vec{\nu}$. The nuisance parameters encompass all parameters that are listed in \reftab{num-input},
as well as the parameters $r_\lambda^{\Gamma,\tilde\Gamma}$ that are specified in \refapp{ff-fit}.

For the nominal integration range we obtain the modes of the marginalized
distributions and their minimal $68\%$ probability intervals as
\begin{equation}
\begin{aligned}
    \langle \mathcal{B}\rangle                  & =  \big(4.5 \pm 1.2\big) \cdot 10^{-7}\,,\\
    \langle A_\text{FB}^\ell\rangle             & = -0.29 \pm 0.05\,,\\
    \langle A_\text{FB}^\Lambda\rangle          & = -0.26 \pm 0.03\,,\\
    \langle A_\text{FB}^{\ell\Lambda}\rangle    & = +0.13^{+0.02}_{-0.03}\,,\\
    \langle F_0\rangle                          & = +0.4 \pm 0.1\,.
\end{aligned}
\end{equation}

In addition, we compare our result for the somewhat larger $q^2$ bin $14.18\,\GeV^2 \leq q^2 \leq (\mlamB - \mlam)^2$
at $68\%$ probability,
\begin{equation}
    \langle \mathcal{B}\rangle                    =  \big(5.3^{+1.5}_{-1.3}\big) \cdot 10^{-7}\,,\\
\end{equation}
with our naive combination of the the experimental measurements in two low-recoil bins \cite{Aaij:2013hna}\footnote{%
We neglect small correlation effects when combining
the experimental measurements of the two low recoil bins.
},
\begin{equation}
    \langle \mathcal{B}\rangle^\text{LHCb}        =  \big(6.2 \pm 2.1\big) \cdot 10^{-7}\,.\\
\end{equation}
We find good agreement between our prediction and the measurement, which is not surprising given the substantial
uncertainties that affect both the prediction of the branching ratio and the experimental measurement of the same.

In light of the present tension between the recent LHCb measurement with some of the SM predictions \cite{Aaij:2013qta}
-- see also \cite{Jager:2012uw} for more compatible predictions -- we further investigate the BSM reach of the angular observables.
For this, we carry out a numerical study of the observables $X_{1,2}$ as previously
defined in \refeq{def-x1} and \refeq{def-x1-x2}.
Similar to observables in the decay $\bar{B}\to \bar{K}^{*}(\to \bar{K}\pi)\ell^+\ell^-$, the observables $X_{1,2}$
are constructed so that at small $q^2$ the leading form factor $\xi_\Lambda$ cancels. In addition, they are
sensitive toward BSM effects, specifically right-handed currents. At large $q^2$, $X_1$ remains insensitive to the
hadronic form factors, while $X_2$ develops a small form factor dependence in BSM models with right-handed currents.
For illustration, we produce SM estimates that
are compared to estimates within two benchmark scenarios BM$1$ and BM$2$. We define the latter scenarios as
\begin{equation}
\begin{aligned}
    \wilson[\mathrm{BM1}]{9}  & = \wilson[\mathrm{SM}]{9} - 1\,, &
    \wilson[\mathrm{BM1}]{9'} & = 1\,,
\end{aligned}
\end{equation}
and
\begin{equation}
\begin{aligned}
    \wilson[\mathrm{BM2}]{7}  & = +0.15\,, & \wilson[\mathrm{BM2}]{9}  & =  0.0\,, & \wilson[\mathrm{BM2}]{10}  & = -1.0\,, \\
    \wilson[\mathrm{BM2}]{7'} & = +0.40\,, & \wilson[\mathrm{BM2}]{9'} & = -4.0\,, & \wilson[\mathrm{BM2}]{10'} & = -4.5\,,
\end{aligned}
\end{equation}
while the rest of the Wilson coefficients remain as in the SM, respectively. Our motivation for these scenarios stems from
a comprehensive analysis of available $b\to s$-FCNC decays \cite{Beaujean:2013soa}; see also \cite{Altmannshofer:2013foa,Descotes-Genon:2013wba} for further works.
Specifically, scenario BM1 corresponds to the best-fit point in a constrained fit of only $\wilson{9,9'}$, as can be seen
in figure 4 of \cite{Beaujean:2013soa}. Scenario BM2 corresponds to solution $D'$ in a fit to the SM and chirality-flipped Wilson coefficients,
as can be seen in figure 3 of \cite{Beaujean:2013soa}. Both scenarios can explain the present tension at $68\%$ probability, since they follow
paths in the parameter space of Wilson coefficients that -- for instance -- leave the bilinears $\rho_1^\pm$ and $\rho_2$ constant.
However, both scenarios yield different results for the bilinears $\rho_3^\pm$ and $\rho_4$, which are presently unconstrained.
We thus expect to be able to discriminate them based on sufficiently precise measurements of $\Lambda_b\to \Lambda(\to N\pi)\ell^+\ell^-$ observables.\\

For the low recoil bin $15\,\GeV^2 \leq q^2 \leq (\mlamB-\mlam)^2$ we obtain
\begin{align}
    \langle X_1\rangle^\text{SM} & = +1.54^{+0.03}_{-0.04}\,, & \langle X_1\rangle^\text{BM$1$} & = +1.55^{+0.05}_{-0.04}\,, & \langle X_1\rangle^\text{BM$2$} & = -1.67^{+0.05}_{-0.05}\,, \\
    \langle X_2\rangle^\text{SM} & = +0.63^{+0.01}_{-0.01}\,, & \langle X_2\rangle^\text{BM$1$} & = +0.60^{+0.02}_{-0.03}\,, & \langle X_2\rangle^\text{BM$2$} & = -0.60^{+0.02}_{-0.02}\,.
\end{align}
For the commonly used large recoil bin $1\,\GeV^2 \leq q^2 \leq 6\,\GeV^2$ we find
\begin{align}
    \langle X_1\rangle^\text{SM} & = +0.08^{+0.12}_{-0.09}\,, & \langle X_1\rangle^\text{BM$1$} & = -0.49^{+0.07}_{-0.08}\,, & \langle X_1\rangle^\text{BM$2$} & = +0.35^{+0.10}_{-0.15}\,, \\
    \langle X_2\rangle^\text{SM} & = +0.17^{+0.04}_{-0.17}\,, & \langle X_2\rangle^\text{BM$1$} & = -0.22^{+0.03}_{-0.03}\,. & \langle X_2\rangle^\text{BM$2$} & = +0.33^{+0.09}_{-0.31}\,.
\end{align}
In both cases, the estimation of uncertainties is the same as for the simple observables above. However, we emphasize that the uncertainty estimate
for the large recoil region is not very rigorous because numerically important
contributions from hard spectator interactions are not yet known.

Our numerical estimates clearly show that the observables $X_{1,2}$ are capable to distinguish between the benchmark models BM$1$, BM$2$
and the SM as measurements both at large and at low hadronic recoil are taken into account. In particular, a distinction between
the SM and BM$1$ can be made using precise measurements of $X_{1,2}$ at small $q^2$. Similarly, the SM and BM$2$ can be distinguished
using large-$q^2$ measurements.

\begin{table}
    \begin{center}
    \begin{tabular}{c||ccc}
        \tabvspbot
        parameter             & value and 68\% interval & unit & source\\
        \hline
        \hline \tabvsptop \tabvspbot
        $\lambda$             & $0.2253\pm 0.0006$      &      & \cite{Bona:2006ah}\\
        \hline \tabvsptop \tabvspbot
        $A$                   & $0.806 \pm 0.020$       &      & \cite{Bona:2006ah}\\
        \hline \tabvsptop \tabvspbot
        $\overline\rho$       & $0.132 \pm 0.049$       &      & \cite{Bona:2006ah}\\
        \hline \tabvsptop \tabvspbot
        $\overline\eta$       & $0.369 \pm 0.050$       &      & \cite{Bona:2006ah}\\
        \hline \tabvsptop \tabvspbot
        $\alpha$              & $0.642 \pm 0.013$       &      & \cite{Beringer:1900zz}\\
        \hline \tabvsptop \tabvspbot
        $\overline{m_c}(m_c)$ & $1.275 \pm 0.025$       & \GeV & \cite{Beringer:1900zz}\\
        \hline \tabvsptop \tabvspbot
        $\overline{m_b}(m_b)$ & $4.18  \pm 0.03$        & \GeV & \cite{Beringer:1900zz}\\
    \end{tabular}
\end{center}
\caption{Summary of the prior distributions for the nuisance parameters $\vec\nu$ (except form factor ratios)
    that enter the observables in addition to the form factor parameters $\vec x$.
    \label{tab:num-input}}
\end{table}

\section{Summary and Outlook}

In the present article we have investigated the phenomenological potential of 
the rare decay $\Lambda_b\to \Lambda\ell^+\ell^-$ with a subsequent, self-analyzing 
$\Lambda \to N \pi$ transition. From the kinematics of the primary and 
secondary decay we have worked out the fully differential decay width that follows
from the Standard Model (SM) operator basis for radiative $b \to s$ transitions and its
chirality-flipped counterpart which may be relevant for physics beyond the SM.
Similar to the corresponding mesonic decay, $B\to (K^* \to K \pi) \ell^+\ell^-$, the
differential decay width can be expressed in terms of 10 angular observables.
In the (naive) factorization approximation, these can be conveniently expressed
in terms of short-distance Wilson coefficients and hadronic transition form factors
in the helicity or transversity basis.

Exploiting the simplifications that arise in the heavy $b$-quark mass limit
-- noteably the form-factor relations that arise 
in the framework of heavy-quark effective theory for low recoil, or
soft-collinear effective theory for large recoil --  we have discussed
the phenomenological consequences for some interesting 
observables: In the SM the fraction of transverse dilepton polarization,
and various forward-backward asymmetries in the leptonic or
baryonic variables show a characteristic dependence on the leptonic
invariant mass $q^2$ which can be confronted with experimental data.

Numerical predictions for these observables have been obtained
on the basis of a careful statistical analysis
of the presently available estimates of hadronic input parameters 
and their (correlated) uncertainties.
We have also identified a number of ratios of angular observables
where either the short-distance Wilson coefficients or the 
long-distance form factors drop out to first approximation.
In particular, as a consequence of the parity-violating nature of
the secondary decay, we have found that the angular analysis of
the $\Lambda_b \to \Lambda (\to N\pi) \ell^+\ell^-$ decay is sensitive
to combinations of Wilson coefficients that cannot be directly tested
in $B \to K^*(\to K\pi) \ell^+\ell^-$ decays.
Future experimental information on these ratios can thus be used
to complement the on-going search for new physics from rare radiative
$b \to s$ transitions.

Compared to the mesonic counterpart decays, $B \to K^{(*)}\ell^+\ell^-$,
both the theoretical and experimental situation is not yet competitive:
Detailed experimental information, in particular for the large-recoil region,
is still lacking at the moment; from the theoretical side, a systematic
analysis of non-factorizable hadronic effects (i.e.\ not form-factor like)
is still missing. Both issues are expected to be (at least partially) 
solved in the future,
and the decay $\Lambda_b \to \Lambda (\to N\pi)\ell^+\ell^-$
can thus play an important role in the flavour-physics program at the 
Large Hadron Collider.

\section*{Acknowledgements}

D.v.D would like to thank Patrick Jussel and Michal Kreps for discussions regarding the
experimental analysis of $\Lambda_b \to \Lambda\ell^+\ell^-$ decays,
and Imkong Sentitemsu Imsong and Thomas Mannel for discussions on
$\Lambda_b\to \Lambda$ form factors.
We would also like to thank the authors of \cite{Das:2014sra} for sharing their results with us
prior to publication.\\
We are grateful to Stefan Meinel for pointing out a typo in \refeq{angular-distribution}.
This work is supported in parts by the Bundesministerium f\"ur Bildung und Forschung (BMBF),
and by the Deutsche Forschungsgemeinschaft (DFG) within Research Unit FOR 1873
(``Quark Flavour Physics and Effective Field Theories'').

\appendix

\section{Corrections to HQET Form Factor Relations}
\label{sec:app:hqet-beyond-leading-power}

In the low-recoil region, the $\Lambda_b \to \Lambda$ form factors 
are related by HQET spin symmetries in the heavy-quark limit.
We follow the analysis in \cite{Grinstein:2004vb} and take into account
sub-leading terms in $\alpha_s$ and $1/m_b$ appearing in the matching of QCD currents
to HQET. For the vector and axial-vector currents
this amounts to
\begin{align}
    \bar{s} \gamma^\mu b
    & =   C_0^{(v)} \, \bar{s}\gamma^\mu h_v +
          C_1^{(v)} v^\mu \, \bar{s} h_v 
          + \frac{1}{2m_b} \, \bar{s}\gamma^\mu i\slashed{D}_\perp  h_v 
          + \ldots \,,
          \cr
      \bar{s} \gamma^\mu \gamma_5 b
    & =    C_0^{(v)} \, \bar{s}\gamma^\mu \gamma_5 h_v 
          - C_1^{(v)} v^\mu \, \bar{s} \gamma_5 h_v 
          - \frac{1}{2m_b} \, \bar{s}\gamma^\mu i\slashed{D}_\perp \gamma_5 h_v 
          + \ldots\,,
\end{align}    
and for the tensor and pseudotensor currents one has    
 \begin{align}   
    \bar{s} i\sigma^{\mu\nu} q_\nu (\gamma_5) b
    & = C_0^{(t)} \, \bar{s} i \sigma^{\mu\nu} q_\nu (\gamma_5) h_v 
    \pm \frac{1}{2m_b} \, \bar{s}\sigma^{\mu\nu} q_\nu i\slashed{D}_\perp (\gamma_5) h_v 
    + \ldots
        \end{align}
Here the leading-power matching coefficients at NLO read
\begin{equation}
\begin{aligned}
    C_0^{(v)}(\mu) & = 1 - \frac{\alpha_s C_F}{4\pi}\left(3 \ln \left(\frac{\mu}{m_b}\right) + 4\right) + \order{\alpha_s^2}\,,\\
    C_1^{(v)}(\mu) & = \frac{\alpha_s C_F}{2\pi} + \order{\alpha_s^2}\,,\\
    C_0^{(t)}(\mu) & = 1 - \frac{\alpha_s C_F}{4\pi}\left(5 \ln \left(\frac{\mu}{m_b}\right) + 4\right) + \order{\alpha_s^2}\,.
\end{aligned}
\end{equation}
The hadronic matrix elements of these currents can then parametrized in
terms of leading and sub-leading Isgur-Wise functions, denoted as 
 $\xi_n \equiv \xi_n(v\cdot k)$ and
$\chi_m \equiv \chi_m(v \cdot k)$, respectively:
\begin{align}
    & \langle \Lambda(k, s_\Lambda)|\bar{s} \gamma^\mu (\gamma_5) b |\Lambda_b(p = \mlamB v, s_{\Lambda_b})\rangle\cr
    \simeq \  & C_0^{(v)} \sum_{n=1,2} \xi_n \,
    \bar{u}_\Lambda(k, s_\Lambda)\Gamma_n \gamma^\mu (\gamma_5) u_{\Lambda_b}(v, s_{\Lambda_b})\cr
    &  \pm C_1^{(v)} \sum_{n=1,2} \xi_n v^\mu \,
    \bar{u}_\Lambda(k, s_\Lambda)\Gamma_n (\gamma_5) u_{\Lambda_b}(v, s_{\Lambda_b})\cr
    &  \pm \sum_{m} \frac{\chi_m}{2 m_b} \,
    \bar{u}_\Lambda(k, s_\Lambda)\hat{\Gamma}_m \gamma^\mu (\gamma_5)\tilde\Gamma_m u_{\Lambda_b}(v, s_{\Lambda_b})
\label{eq:sl-matching-hqet-vector}
\end{align}
and
\begin{align}
    & \langle \Lambda(k, s_\Lambda)|\bar{s} i\sigma^{\mu\nu} q_\nu (\gamma_5) b |\Lambda_b(p = \mlamB v, s_{\Lambda_b})\rangle\cr
    \simeq \ & C_0^{(t)} \sum_{n=1,2} \xi_n \, 
    \bar{u}_\Lambda(k, s_\Lambda)\Gamma_n i \sigma^{\mu\nu} q_\nu (\gamma_5) u_{\Lambda_b}(v, s_{\Lambda_b})\cr
    & \pm \sum_{m} \frac{\chi_m}{2 m_b} \,
    \bar{u}_\Lambda(k, s_\Lambda)\hat{\Gamma}_m i\sigma^{\mu\nu} q_\nu (\gamma_5)\tilde\Gamma_m 
    u_{\Lambda_b}(v, s_{\Lambda_b})    \,.
\label{eq:sl-matching-hqet-tensor}
\end{align}
Here, the independent Dirac structures are given by
\begin{equation}
    \Gamma_1  = 1\,, \qquad 
    \Gamma_2  = \slashed{v}\,,
\end{equation}
for the leading-power terms, and
\begin{equation}
\begin{aligned}
    \hat\Gamma_1 & = \mlam \gamma_\mu    \,,          \qquad &    \tilde\Gamma_1 & = \gamma^\mu_\perp\,,\cr
    \hat\Gamma_2 & = k_\mu                       \,,  &    \tilde\Gamma_2 & = \gamma^\mu_\perp\,,\cr
    \hat\Gamma_3 & = \mlam \gamma_\mu \gamma_5\,,     &    \tilde\Gamma_3 & = \gamma^\mu_\perp \gamma_5\,,\cr
    \hat\Gamma_4 & = k_\mu \gamma_5             \,,   &    \tilde\Gamma_4 & = \gamma^\mu_\perp \gamma_5\,,\cr
    \hat\Gamma_5 & = \frac{i}{2}\gamma^\mu\gamma_5\,, &    \tilde\Gamma_5 & = \gamma_{\perp}^\nu v^\alpha k^\beta \eps_{\mu\nu\alpha\beta}\,,\cr
    \hat\Gamma_6 & = \frac{i}{2}\gamma^\mu         \,, &    \tilde\Gamma_6 & = \gamma_{\perp}^\nu \gamma_5 v^\alpha k^\beta \eps_{\mu\nu\alpha\beta}\,,\\
\end{aligned}
\end{equation}
for the terms at subleading power.
For the physical form factors this translates into
{%\small
\begin{align}
    f_\perp^{V,A}    & = C_0^{(v)} \big(\xi_1 \mp \xi_2\big) - \frac{\mlam (\chi_1 + \chi_3)}{2 m_b} \mp \frac{s_\pm (\chi_2 + \chi_4)}{4 m_b \mlamB}\\
    f_0^{V,A}        & = \left(C_0^{(v)} + \frac{C_1^{(v)} s_\pm}{2 \mlamB (\mlamB \pm \mlam)}\right)\xi_1 
                 \mp \left(C_0^{(v)} - \frac{(2 C_0^{(v)} + C_1^{(v)}) s_\pm}{2 \mlamB (\mlamB \pm \mlam)}\right) \xi_2\cr
                 & \quad - \frac{\mlam}{2 m_b}\left(1 + \frac{s_\pm}{\mlamB(\mlamB \pm \mlam)}\right) (\chi_1 + \chi_3) \cr
                 & \quad \pm \frac{s_\pm}{4 m_b \mlamB}\frac{\mlamB \mp \mlam}{\mlamB \pm \mlam}\big[(\chi_2 + \chi_4) \pm (\chi_5 + \chi_6)\big]\,,
\end{align}
}%
and %for the tensor form factors
\begin{align}
     f_\perp^{T(5)}   
                 & = C_0^{(t)}\left((\xi_1 \mp \xi_2) \pm \frac{s_\pm}{\mlamB (\mlamB \pm \mlam)}\xi_2\right)\cr
                 & \quad + \frac{\mlam}{2 m_b}\left(1 - \frac{s_\pm}{\mlamB (\mlamB \pm \mlam)}\right)\big(\chi_1 - \chi_3\big)\cr
                 & \quad \pm \frac{\mlamB \mp \mlam}{\mlamB \pm \mlam} \frac{s_\pm}{4 m_b\mlamB} \big(\chi_2 - \chi_4\big)\,,\\
    f_0^{T(5)}
                 & = C_0^{(t)} (\xi_1 \mp \xi_2) + \frac{\mlam}{2m_b} (\chi_1 - \chi_3)
                   \mp \frac{s_\pm}{4m_b\mlamB} \big[(\chi_2 - \chi_4) \pm (\chi_5 - \chi_6)\big]\,.
\end{align}

\section{Form Factor Parametrisation}
\label{sec:app:ff-param}

The form factors for $\Lambda_b\to \Lambda$ transitions can be treated in a similar 
way as those for e.g.\ $B\to K^{(*)}$ transitions. 
To this end, one considers the analytic continuation into the complex
$t$ plane, where $q^2 = \Re{t}$. For $0 \leq q^2 \leq t_-$ (semileptonic domain)
the form factors describe the semileptonic decay region, $\Lambda_b\to \Lambda$.
For $q^2 \geq t_+$ they describe the pair-production process $|0\rangle \to \overline{\Lambda}_b \Lambda$
(pair production domain). Here $t_\pm = (\mlamB \pm \mlam)^2$, and $|0\rangle$ denotes the hadronic vacuum.

Inbetween the two domains (i.e., for $t_- \leq q^2 \leq t_+$) exists a region where the unphysical process
$\Lambda_b \to \Lambda H$ contributes. Here the \emph{low-lying resonance}
$H$ with mass $m_H$ denotes any hadron with one bottom quark and one strange anti-quark
that fulfills
\begin{equation}
    t_- < m_H^2 < t_+\,.
\end{equation}
A summary of these resonances and their spin-parity quantum numbers 
is given in \reftab{resonances}.
\begin{table}[t!]
    \begin{center}
    \begin{tabular}{c||cccc}
        \tabvspbot
        $J^P$          & $0^-$   & $0^+$   & $1^-$   & $1^+$\\
        \hline
        \hline \tabvsptop \tabvspbot
        mass [GeV$^2$] & $5.367$ & unknown & $5.415$ & $5.829$
    \end{tabular}
    \end{center}
    \caption{List of low-lying $B_s$ resonances for the transition form factor $\Lambda_b\to \Lambda$ as taken
        from \cite{Beringer:1900zz}
    \label{tab:resonances}}
\end{table}
The contribution of the corresponding poles in the complex plane 
can be included in the parametrization of the form factors as follows
(see e.g.\ \cite{Bourrely:2008za,Khodjamirian:2011jp}),
\begin{align}
    \label{eq:ff-param}
    f_\lambda^V(q^2) & = \frac{f_\lambda^V(0)}{P(q^2, m_{B_s^*}(1^-))}\left[1 + b^{V,\lambda}_1 \big(z(q^2, t_0) - z(0, t_0)\big) + \dots\right]\,,\\
    f_\lambda^A(q^2) & = \frac{f_\lambda^A(0)}{P(q^2, m_{B_s^*}(1^+))}\left[1 + b^{A,\lambda}_1 \big(z(q^2, t_0) - z(0, t_0)\big) + \dots\right]\,.
\end{align}
for $\lambda = 0,\perp$. Here the resonance factor is defined as
\begin{equation}
    P(t, m_H^2) = 1 - \frac{t}{m_H^2}\,,
\end{equation}
and the remaining $q^2$-dependence is obtained from a Taylor series in
the variable 
\begin{equation}
    z(t, t_0) \equiv \frac{\sqrt{t_+ - t} - \sqrt{t_+ - t_0}}{\sqrt{t_+ - t} + \sqrt{t_+ - t_0}}\,,
\end{equation}
which corresponds to a conformal mapping of the complex $t$ plane onto the unit disc $|z| \leq 1$.
Given the currently available numerical precision for the form factors, 
it is sufficient to truncate the expansion after the first order.
For the auxiliary parameter $t_0$ we choose $t_0 = 12~\GeV^2$.

\section{Statistical Analysis of the Transition Form Factors}
\label{sec:app:ff-fit}

This appendix is dedicated to details of our fit of the hadronic transition form factors.
Specifically, we undertake the following steps.
\begin{itemize}
    \item[(a)] The available lattice results \cite{Detmold:2012vy} provide information on the HQET form factors 
        $\xi_{1}$ and $\xi_{2}$ in the interval $q^2 \in [13.5,20.5]~\GeV^2$ (low recoil energy). This includes correlations between 
        form-factor values at different values of $q^2$ which -- non-surprisingly -- are sizeable (up to 
        $\simeq 100$\% for form factors related to the same current at adjacent points). 
        For a stable numerical analysis we therefore only consider the two $q^2$ values at the very boundaries
        of the simulated range which are expected to have minimal correlation.

        To translate this into estimates for the helicity form factors $f_{\perp,0}^{V,A}(q^2)$,
        we allow for corrections to the HQET limit from subleading form factors $\chi_{n=1\ldots 6}$, see \refapp{hqet-beyond-leading-power}.
        With no detailed information on the latter -- following the principle of maximum entropy \cite{Jaynes:2003} -- 
        we use Gaussian priors, i.e.\ a normal distribution centered around zero with variance $\sigma= 1$ from naive 
        power-counting. Notice that by construction this leads to prior distributions with flat $q^2$-dependence.

        The present lattice analysis do not provide correlation information 
        between the two individual HQET form factors. Given the large correlations between
        individual $q^2$ points, we expect these correlations to be sizeable, too.
        As a consequence, we cannot completely determine the correlation
        between the data points for, say, the physical vector form factor.
        Notice that the correlation is further enhanced by the flat priors
        for the subleading contributions $\chi_n$. 
        This issue can be solved as soon as improved lattice results for the physical form factors
        beyond the HQET limit will be available.

        The result of this part of the analysis is summarized in \reftab{ff-constraints:lattice}.
        However, we find that the estimates correlations in excess of $97\%$ lead to a degeneracy
        of the parameters in our fits. Hereafter, our incomplete estimates of the
        correlation between $q^2$ points are therefore disregarded.
\end{itemize}

\begin{table}[t!bph]
    \begin{center}
    \begin{tabular}{c||c|c||c|c}
        \tabvspbot
        $q^2$                      & $f_\perp^V$     & $f_0^V$        & $f_\perp^A$    & $f_0^A$\\
        \hline
        \hline \tabvsptop \tabvspbot
        $13.5~\GeV^2$              & $0.73\pm 0.20$  & $0.72\pm 0.21$ & $0.48\pm 0.19$ & $0.48 \pm 0.19$\\
        \hline \tabvsptop \tabvspbot
        $20.5~\GeV^2$              & $1.40\pm 0.20$  & $1.39\pm 0.21$ & $0.85\pm 0.19$ & $0.85 \pm 0.19$\\
%        \hline
%        \hline \tabvsptop \tabvspbot
%        $q^2$ correlation          & $0.977$         & $0.981$        & $0.998$        & $0.994$\\
    \end{tabular}
    \end{center}
    \caption{Estimates of mean values and standard deviations for helicity form factors $f_\perp^{V,A}$ and $f_0^{V,A}$,
    based on probability distributions obtained from lattice points \cite{Detmold:2012vy} in the HQET limit
     and Gaussian priors for the subleading IW form factors. See text for details. 
    \label{tab:ff-constraints:lattice}}
\end{table}

\begin{table}[t!!!pbph]
    \begin{center}
    \begin{tabular}{c||c|c||c|c}
        \tabvspbot
        $q^2$                      & $R_\perp^{T,V}$ & $R_0^{T,V}$    & $R_\perp^{T5,A}$ & $R_0^{T5,A}$\\
        \hline
        \hline \tabvsptop \tabvspbot
        $13.5~\GeV^2$              & $1.07\pm 0.45$  & $1.10\pm 0.55$ & $1.13 \pm 0.66$  & $1.12 \pm 0.67$\\
        \hline \tabvsptop \tabvspbot
        $20.5~\GeV^2$              & $1.00\pm 0.19$  & $1.03\pm 0.21$ & $1.02 \pm 0.30$  & $1.03 \pm 0.30$\\
        \hline
        \hline \tabvsptop \tabvspbot
        $q^2$ correlation          & $0.965$         & $0.964$        & $0.961$          & $0.956$\\
        \hline
        \hline \tabvsptop \tabvspbot
        $a_\lambda^{\Gamma,\tilde\Gamma}$
                                   & $1.205$         & $1.235$        & $1.342$          & $1.294$\\
        \hline \tabvsptop \tabvspbot
        $b_\lambda^{\Gamma,\tilde\Gamma}\cdot10^2$
                                   &$-1.00$          &-$1.00$         &$-1.57$           &$-1.29$ \\
        \hline \tabvsptop \tabvspbot
        $c_\lambda^{\Gamma,\tilde\Gamma}$
                                   & $5.01$          & $5.74$         & $4.51$           & $4.61$ \\
        \hline \tabvsptop \tabvspbot
        $d_\lambda^{\Gamma,\tilde\Gamma}$
                                   &$-0.195$         &$-0.231$        &$-0.171$          &$-0.176$\\
        \hline \tabvsptop \tabvspbot
        $\sigma_\lambda^{\Gamma,\tilde\Gamma}$
                                   & $0.19$          & $0.21$         & $0.30$           & $0.30$ \\
    \end{tabular}
    \end{center}
    \caption{Estimates of mean values and standard deviations for ratios of helicity form factors $R_\lambda^{\Gamma,\tilde\Gamma}$,
    based on probability distributions obtained from lattice points \cite{Detmold:2012vy} in the HQET limit
     and Gaussian priors for the subleading IW form factors. See text for details. 
    \label{tab:ff-constraints:lattice2}}
\end{table}

\begin{itemize}
    \item[(b)] 
        As explained in the main text, the form factors for tensor and pseudotensor currents enter the observables in the form of \emph{ratios}
        with the corresponding vector or axial-vector form factors. We therefore derive probability distributions 
        for the ratios ($\lambda={0,\perp}$, $\Gamma =T,T5$, $\tilde\Gamma=V,A$)
        \begin{equation}
            R_\lambda^{\Gamma,\tilde\Gamma} \equiv \frac{f_\lambda^\Gamma}{f_\lambda^{\tilde\Gamma}}\,,
        \end{equation}
        again based on the lattice results \cite{Detmold:2012vy} at the boundaries
        $q^2 = 13.5~\GeV^2$ and $q^2 = 20.5~\GeV^2$. In the heavy-quark limit these
        ratios are unity.
        Deviations are estimated by the same implementation of sub-leading corrections
        as described in (a).

        The result of this part of the analysis is summarized in \reftab{ff-constraints:lattice2}.

        Notice, that the ratios $R_{\lambda}^{\Gamma,\tilde\Gamma}$ are only very poorly constrained towards
        smaller values of $q^2$. More precise inputs on these quantities directly from the lattice
        are desirable.

        The high degree of correlation between $q^2$ points leads us to parametrize these ratios
        through linear interpolation in $q^2$, and one random number $r$ for their uncertainty. We use
        \begin{equation}
            R_\lambda^{\Gamma,\tilde\Gamma}(q^2)
            = a_\lambda^{\Gamma,\tilde\Gamma} + b_\lambda^{\Gamma,\tilde\Gamma} q^2
            + (c_\lambda^{\Gamma,\tilde\Gamma} + d_\lambda^{\Gamma,\tilde\Gamma} q^2) r_\lambda^{\Gamma,\tilde\Gamma}
        \end{equation}
        with fixed parameters $a_\lambda^{\Gamma,\tilde\Gamma}$ through $d_\lambda^{\Gamma,\tilde\Gamma}$ as given
        in \reftab{ff-constraints:lattice2}.
        In the above, $r_\lambda^{\Gamma,\tilde\Gamma} \sim \mathcal{N}(0, \sigma_\lambda^{\Gamma,\tilde\Gamma})$.

    \item[(c)]
        The above results can be continued to the low-$q^2$ region by     
        taking into account information from sum-rule analyses (see e.g.\ \cite{Wang:2008sm,Wang:2009hra,Feldmann:2011xf})
        to determine probability distributions for the helicity form factors 
        $f_\perp^{V,A}(q^2 = 0)$ and $f_0^{V,A}(q^2 = 0)$ at maximal recoil.
        We have 
        \begin{equation}
            f_{\perp,0}^{V(A)}(0) = C_{\perp,0} \, \xi_{\Lambda}(0)  
            \left(1 \pm  \zeta_{\perp,0} \right)\,,
        \end{equation}
        where we describe the soft form factor by a normal distribution
        $\xi_\Lambda(0) \sim \mathcal{N}(\mu = 0.38, \sigma = 0.19)$ where, for concreteness,
        we have used the estimates from a SCET sum rule obtained in \cite{Feldmann:2011xf}.
        We further allow for two independent (ad-hoc) correction factors $\zeta_\lambda$ 
        for the two helicities $\lambda=0,\perp$ whose probability distributions are modelled
        as $\mathcal{N}(\mu = 0, \sigma = 0.25)$.

        The helicity form factors at $q^2 = 0$ are to good approximation multivariate-normally
        distributed. The mean values and standard deviations read
        \begin{equation}
            \label{eq:SSRmu}
            f_\perp^V(0) = f_\perp^A (0) = 0.39 \pm 0.23\,, \qquad 
            f_0^V(0) = f_0^A (0) = 0.38 \pm 0.22\,.
        \end{equation}
        The form factors are strongly correlated, and we find for the correlation matrix
        \begin{equation}
            \label{eq:SSRrho}
            \rho^\text{SSR} = \left(
                \begin{tabular}{l | cccc}
                    & $f_\perp^V$ & $f_\perp^A$ & $f_0^V$ & $f_0^A$ \\
                    \hline 
                    $f_\perp^V$ &    1.00      & 0.56      & 0.77  & 0.78\\
                    $f_\perp^A$ &    0.56      & 1.00      & 0.77  & 0.77\\
                    $f_0^V$ &        0.77      & 0.77      & 1.00  & 0.53\\
                    $f_0^A$ &        0.78      & 0.77      & 0.53  & 1.00\\
                \end{tabular}
            \right)
        \end{equation}
\end{itemize}

Based on the inputs discussed above in points (a) and (c), we can now fit the helicity form factors
for the vector and axial vector currents to our theory estimates. We use the parametrization given in
\refeq{ff-param}, which features two parameters $f_\lambda^\Gamma(0)$ and $b_\lambda^\Gamma$
per form factor. The overall set of fit parameters therefore reads
\begin{equation}
    \vec x \equiv \big(f_0^V(0), f_0^A(0), f_\perp^V(0), f_\perp^A(0), b_0^V, b_0^A, b_\perp^V, b_\perp^A\big)\,.
    \label{eq:ff-fit:params}
\end{equation}
In our fit we use a uniform prior $P_0(\vec x)$ that is supported on
\begin{equation}
\begin{aligned}
      0 & \leq f_\lambda^\Gamma(0) \leq  1\,, &
    -60 & \leq b_\lambda^\Gamma    \leq +30\,,
\end{aligned}
\end{equation}
for all $\lambda = \perp,0$ and $\Gamma=V,A$. The likelihood for the form factor estimates $P(\text{Estimates} | \vec x)$ factorizes into
\begin{equation}
    P(\text{Estimates}|\vec x) = P(\text{HQET}|\vec x) \times P(\text{SSR} | \vec x)\,.
\end{equation}
The HQET components are
\begin{equation}
    P(\text{HQET}|\vec{x})
    = \prod_{\lambda=\perp,0}^{\Gamma=V,A}
    \mathcal{N}_2\big(\vec\mu^{\text{HQET},\Gamma,\lambda}, \Sigma^{\text{HQET},\Gamma,\lambda}; \vec F_\lambda^\Gamma(\vec{x})\big)\,,
\end{equation}
where $\vec\mu^{\text{HQET},\Gamma,\lambda}$ and $\Sigma^{\text{HQET},\Gamma,\lambda}$ can be
read off \reftab{ff-constraints:lattice}, and $\vec{F}_\lambda^\Gamma$ is an abbreviation for
\begin{equation}
    \vec{F}_\lambda^\Gamma = (f_\lambda^\Gamma(13.5\,\GeV^2; \vec x), f_\lambda^\Gamma(20.5\,\GeV^2; \vec x))\,.
\end{equation}
The SCET sum rule component reads
\begin{equation}
    P(\text{SSR} | \vec x) = \mathcal{N}_4\big(\mu^\text{SSR}, \Sigma^\text{SSR}; \vec{F}(\vec x)\big)\,,\\
\end{equation}
where $\mu^\text{SSR}$ and $\Sigma^\text{SSR}$ can be read off \refeq{SSRmu} and \refeq{SSRrho}.
In the above, we denote the form factor parametrization as
\begin{equation}
    \vec{F}(\vec x) = (f_\perp^V(0; \vec x), f_\perp^A(0; \vec x), f_0^V(0; \vec x), f_0^A(0; \vec x))\,.
\end{equation}
As usual the posterior $P(\vec x|\text{Estimates})$ follows from Bayes' theorem as
\begin{equation}
    \label{eq:ff-fit:posterior}
    P(\vec x|\text{Estimates}) = \frac{P(\text{Estimates}|\vec x) P_0(\vec x)}{\int \dd{\vec x} P(\text{Estimates}|\vec x) P_0(\vec x)}\,.
\end{equation}

We find the best-fit point
\begin{multline}
    \vec{x}^* \equiv \argmax P(\vec x|\text{Estimates})\\
    = (0.33, 0.31, 0.34, 0.31, -1.75, -0.52, -1.58, -0.24)\,.
\end{multline}
This is a good fit, with the largest pull value being $0.05\sigma$. We find $\chi^2 = 5.4\cdot10^{-3}$, and with $N_\text{d.o.f.} = 4$ degrees
of freedom (from 12 observations reduced by 8 fit parameters) this yields a p value of $>0.99$.

\section{Details on the Kinematics}
\label{sec:app:kin}

\subsection{$\Lambda_b$ Rest Frame}

We define the momenta of the lepton pair ($q$) and the $\Lambda$-baryon ($k$) 
in the $\Lambda_b$-baryon rest frame (B-RF) as
\begin{align}
    q^\mu\big|_\text{B-RF}  = \left(q^0, 0, 0, -|\vec{q}\,| \right)\,,\qquad 
    k^\mu\big|_\text{B-RF}  = \left( (m_{\Lambda_b} - q^0), 0, 0, +|\vec{q}\,| \right)\,,
\end{align}
i.e.\ the $z$-axis is along the flight direction of the $\Lambda$, and
\begin{align}
        q^0\big|_\text{B-RF} & = \frac{m_{\Lambda_b}^2 - m_\Lambda^2 + q^2}{2 m_{\Lambda_b}}\,, &
|\vec{q}\,|\big|_\text{B-RF} & = \frac{\sqrt{\lambda(m_{\Lambda_b}^2, m_{\Lambda}^2, q^2)}}{2 m_{\Lambda_b}}\,.
\end{align}

\subsection{Dilepton Rest Frame}

The lepton angle $\theta_\ell$ is defined 
as the angle between the $\ell^-$ direction of flight and 
the $z$-axis in the dilepton rest frame (2$\ell$-RF) 
with the dilepton system decaying in the $x$-$z$--plane,
\begin{align}
q_1^\mu\big|_\text{2$\ell$-RF}      & = (E_\ell, -\qrf \sin\theta_\ell, 0, -\qrf\cos\theta_\ell)\,, \cr
q_2^\mu\big|_\text{2$\ell$-RF}      & = (E_\ell, +\qrf \sin\theta_\ell, 0, +\qrf\cos\theta_\ell)\,,
\end{align}
with
\begin{align}
\qrf\big|_\text{2$\ell$-RF}         &
= \frac{\beta_\ell}{2} \, \sqrt{q^2}\,, &
E_\ell\big|_\text{2$\ell$-RF}       & = \frac{\sqrt{q^2}}{2}\,,
& \beta_\ell                   & = \sqrt{1 - \frac{4m_\ell^2}{q^2}} \,.
\end{align}
This implies
\begin{align}
\bar{q}^\mu\big|_\text{2$\ell$-RF}  & = \sqrt{q^2} \, (0, -\beta_\ell\sin\theta_\ell, 0, -\beta_\ell\cos\theta_\ell)\,.
\end{align}

The various Lorentz scalars that can be built from 
the individual lepton momenta and the dilepton polarization vectors 
can then be written as
\begin{align}
    q_{1} \cdot \eps^{(*)}(0)     & =  - \qrf \, \cos\theta_\ell \,, & 
    q_1 \cdot \eps^{(*)}(\pm)   & = \pm \frac{1}{\sqrt{2}} \, \qrf \, \sin\theta_\ell \,,
    \cr   
   q_2 \cdot \eps^{(*)}(0)     & = +\qrf \, \cos\theta_\ell \,, & 
   q_2 \cdot \eps^{(*)}(\pm)   & = \mp\frac{1}{\sqrt{2}} \, \qrf \, \sin\theta_\ell \,,
\end{align}
and
\begin{align}
% Epsilon Tensors
    \eps_{\mu\nu\rho\sigma} \, \eps(0)^\mu \eps(\pm)^{\nu} q_1^\rho q_2^\sigma
                                & = + \frac{i}{\sqrt{2}} \, \sqrt{q^2} \, \qrf \, \sin\theta_\ell \,, \cr
    \eps_{\mu\nu\rho\sigma} \, \eps(+)^\mu \eps(-)^{\nu} q_1^\rho q_2^\sigma
                                & = + i \, \sqrt{q^2} \, \qrf \, \cos\theta_\ell \,, 
\end{align}
and
\begin{align}
    \eps_{\mu\nu\rho\sigma} \, \eps(+)^\mu \eps(-)^\nu \eps(0)^\rho q_{1,2}^\sigma
                                & = + i \, \frac{\sqrt{q^2}}{2} \,.
\end{align}

\subsection{$N\pi$ System}

The $N\pi$ system is characterized through its invariant mass $k^2 = m_{\Lambda}^2$, 
the angle $\theta_\Lambda$ between the $N$-direction of flight
and the $z$-axis in the $N\pi$ rest frame ($N\pi$-RF),
and an azimuthal angle $\phi$  between the decay plane of the $N\pi$ system and 
the dilepton decay plane,
\begin{align}
    k_1^\mu\big|_\text{$N\pi$-RF}      & = (E_N,   -\krf \sin\theta_\Lambda\cos\phi, -\krf \sin\theta_\Lambda\sin\phi, +\krf \cos\theta_\Lambda)\,,\cr
    k_2^\mu\big|_\text{$N\pi$-RF}      & = (E_\pi, +\krf \sin\theta_\Lambda\cos\phi, +\krf \sin\theta_\Lambda\sin\phi, -\krf \cos\theta_\Lambda)\,,
\end{align}
with
\begin{align}
    \krf           & = \frac{\sqrt{\lambda(k^2, m_N^2, m_\pi^2)}}{2 \sqrt{k^2}} \equiv \frac{\beta_{N\pi}}{2} \, \sqrt{k^2}\,, &
    \beta_{N\pi}  & = \sqrt{\lambda(1, m_N^2/k^2, m_\pi^2/k^2)}\,,\cr
    E_N   & = \sqrt{m_N^2 + \frac{\beta_{N\pi}^2 \, k^2}{4}}\,, &
    E_\pi & = \sqrt{m_\pi^2 + \frac{\beta_{N\pi}^2 \, k^2}{4}}\,.
\end{align}
Our convention for the azimuthal angle $\phi$ 
is consistent with that of Kr\"uger/Matias \cite{Kruger:2005ep}.

\section{Explicit Spinor Representations}
\label{sec:app:spinors}

In order to calculate the various helicity amplitudes, 
we have used explicit expressions for the Dirac spinors that characterize
baryons with a given momentum 
\begin{equation}
    p^\mu = (p^0, |\vec{p}|\sin\theta\cos\phi, |\vec{p}|\sin\theta\sin\phi, |\vec{p}|\cos\theta),
\end{equation}
and spin orientation $s = \pm 1/2$ (defined in their respective rest frames), see e.g.\ \cite{Haber:1994pe}:
\begin{align}
    u(p, s=+1/2) & = \frac{1}{\sqrt{2 (p^0 + m)}} \left[\begin{matrix}
    +(p^0 + m - |\vec{p}|) & \cos(\theta/2) & \\
    +(p^0 + m - |\vec{p}|) & \sin(\theta/2) & \exp(+i \phi)\\
    +(p^0 + m + |\vec{p}|) & \cos(\theta/2) & \\
    +(p^0 + m + |\vec{p}|) & \sin(\theta/2) & \exp(+i \phi)
    \end{matrix}\right]\nonumber \\[0.2em]
    u(p, s=-1/2) & = \frac{1}{\sqrt{2 (p^0 + m)}} \left[\begin{matrix}
    -(p^0 + m + |\vec{p}|) & \sin(\theta/2) & \exp(-i\phi)\\
    +(p^0 + m + |\vec{p}|) & \cos(\theta/2) & \\
    -(p^0 + m - |\vec{p}|) & \sin(\theta/2) & \exp(-i\phi)\\
    +(p^0 + m - |\vec{p}|) & \cos(\theta/2) &
    \end{matrix}\right]\,.
\end{align}
For the helicity amplitudes characterizing $\Lambda_b \to \Lambda$ transitions
with scalar, pseudoscalar, vector 
or axialvector currents, we then obtain
\begin{align}
\bar{u}(k, \pm 1/2) \, u(p, \pm 1/2) & = \sqrt{s_+} \,, &
 \bar{u}(k, \pm 1/2) \, u(p, \mp 1/2) & = 0 \,,  \cr 
 \bar{u}(k, \pm 1/2) \, \gamma_5 \, u(p, \pm 1/2)    
 & = \mp \sqrt{s_-}\,, &
  \bar{u}(k, \pm 1/2) \, \gamma_5\, u(p, \mp 1/2)   & = 0 \,,
\end{align}
and
\begin{align}
 \bar{u}(k, \pm 1/2) \, \gamma^\mu \, u(p, \pm 1/2) 
   & = \left( \sqrt{s_+},0,0,\sqrt{s_-}\right)\,, 
    \cr 
 \bar{u}(k, \pm 1/2) \, \gamma^\mu \gamma_5 \, u(p, \pm 1/2)    
    & = \pm \left(
        \sqrt{s_-} , 0 , 0 , \sqrt{s_+}
\right) \,,
\end{align}
and    
 \begin{align}   
    \bar{u}(k, \pm 1/2) \, \gamma^\mu \, u(p, \mp 1/2)  & 
    = \sqrt{2 s_-} \, \eps^\mu(\pm) \,, \cr 
     \bar{u}(k, \pm 1/2) \, \gamma^\mu \gamma_5 \, u(p, \mp 1/2)  & 
    = \mp \sqrt{2 s_+} \, \eps^\mu(\mp) \,.
\end{align}
Here the kinematic functions $s_\pm$ have been defined in \refeq{spm}.

%% Bibliography

\bibliographystyle{JHEP}
\bibliography{references.bib}

\providecommand{\href}[2]{#2}\begingroup\raggedright\begin{thebibliography}{10}

\bibitem{Buchalla:2008jp}
M.~Artuso, D.~Asner, P.~Ball, E.~Baracchini, G.~Bell, et~al., {\it {$B$, $D$
  and $K$ decays}},  {\em Eur.Phys.J.} {\bf C57} (2008) 309--492,
  [\href{http://xxx.lanl.gov/abs/0801.1833}{{\tt arXiv:0801.1833}}].

\bibitem{Antonelli:2009ws}
M.~Antonelli, D.~M. Asner, D.~A. Bauer, T.~G. Becher, M.~Beneke, et~al., {\it
  {Flavor Physics in the Quark Sector}},  {\em Phys.Rept.} {\bf 494} (2010)
  197--414, [\href{http://xxx.lanl.gov/abs/0907.5386}{{\tt arXiv:0907.5386}}].

\bibitem{Bediaga:2012py}
{\bf LHCb Collaboration} Collaboration, R.~Aaij et~al., {\it {Implications of
  LHCb measurements and future prospects}},  {\em Eur.Phys.J.} {\bf C73} (2013)
  2373, [\href{http://xxx.lanl.gov/abs/1208.3355}{{\tt arXiv:1208.3355}}].

\bibitem{Bevan:2014iga}
{\bf BaBar Collaboration, Belle Collaboration} Collaboration, A.~Bevan et~al.,
  {\it {The Physics of the B Factories}},
  \href{http://xxx.lanl.gov/abs/1406.6311}{{\tt arXiv:1406.6311}}.

\bibitem{Aaij:2013hna}
{\bf LHCb collaboration} Collaboration, R.~Aaij et~al., {\it {Measurement of
  the differential branching fraction of the decay
  $\Lambda_b^0\rightarrow\Lambda\mu^+\mu^-$}},  {\em Phys.Lett.} {\bf B725}
  (2013) 25, [\href{http://xxx.lanl.gov/abs/1306.2577}{{\tt arXiv:1306.2577}}].

\bibitem{Hiller:2001zj}
G.~Hiller and A.~Kagan, {\it {Probing for new physics in polarized $\Lambda_b$
  decays at the $Z$}},  {\em Phys.Rev.} {\bf D65} (2002) 074038,
  [\href{http://xxx.lanl.gov/abs/hep-ph/0108074}{{\tt hep-ph/0108074}}].

\bibitem{Chen:2001zc}
C.-H. Chen and C.~Geng, {\it {Baryonic rare decays of $\Lambda_b \to \Lambda
  \ell^+ \ell^-$}},  {\em Phys.Rev.} {\bf D64} (2001) 074001,
  [\href{http://xxx.lanl.gov/abs/hep-ph/0106193}{{\tt hep-ph/0106193}}].

\bibitem{Aliev:2010uy}
T.~Aliev, K.~Azizi, and M.~Savci, {\it {Analysis of the $\Lambda_{b}\rightarrow
  \Lambda \ell^+\ell^- $ decay in QCD}},  {\em Phys.Rev.} {\bf D81} (2010)
  056006, [\href{http://xxx.lanl.gov/abs/1001.0227}{{\tt arXiv:1001.0227}}].

\bibitem{Azizi:2013eta}
K.~Azizi, S.~Kartal, A.~Olgun, and Z.~Tavukoglu, {\it {Analysis of the
  semileptonic $\Lambda_b\rightarrow \Lambda \ell^+ \ell^-$ transition in the
  topcolor-assisted technicolor model}},  {\em Phys.Rev.} {\bf D88} (2013),
  no.~7 075007, [\href{http://xxx.lanl.gov/abs/1307.3101}{{\tt
  arXiv:1307.3101}}].

\bibitem{Gutsche:2013pp}
T.~Gutsche, M.~A. Ivanov, J.~G. K{\"o}rner, V.~E. Lyubovitskij, and
  P.~Santorelli, {\it {Rare baryon decays $\Lambda_b \to \Lambda \ell^+ \ell^-$
  ($\ell=e, \mu, \tau$) and $\Lambda_b \to \Lambda \gamma$ : differential and
  total rates, lepton- and hadron-side forward-backward asymmetries}},
  \href{http://xxx.lanl.gov/abs/1301.3737}{{\tt arXiv:1301.3737}}.

\bibitem{Wang:2011uv}
W.~Wang, {\it {Factorization of Heavy-to-Light Baryonic Transitions in SCET}},
  {\em Phys.Lett.} {\bf B708} (2012) 119--126,
  [\href{http://xxx.lanl.gov/abs/1112.0237}{{\tt arXiv:1112.0237}}].

\bibitem{Detmold:2012vy}
W.~Detmold, C.-J.~D. Lin, S.~Meinel, and M.~Wingate, {\it {$\Lambda_b \to
  \Lambda \ell^+\ell^-$ form factors and differential branching fraction from
  lattice QCD}},  {\em Phys.Rev.} {\bf D87} (2013), no.~7 074502,
  [\href{http://xxx.lanl.gov/abs/1212.4827}{{\tt arXiv:1212.4827}}].

\bibitem{Feldmann:2011xf}
T.~Feldmann and M.~W. Yip, {\it {Form Factors for $\Lambda_b \to \Lambda$
  Transitions in SCET}},  {\em Phys.Rev.} {\bf D85} (2012)
  [\href{http://xxx.lanl.gov/abs/1111.1844}{{\tt arXiv:1111.1844}}].

\bibitem{Ali:2012pn}
A.~Ali, C.~Hambrock, A.~Y. Parkhomenko, and W.~Wang, {\it {Light-Cone
  Distribution Amplitudes of the Ground State Bottom Baryons in HQET}},  {\em
  Eur.Phys.J.} {\bf C73} (2013) 2302,
  [\href{http://xxx.lanl.gov/abs/1212.3280}{{\tt arXiv:1212.3280}}].

\bibitem{Bell:2013tfa}
G.~Bell, T.~Feldmann, Y.-M. Wang, and M.~W.~Y. Yip, {\it {Light-Cone
  Distribution Amplitudes for Heavy-Quark Hadrons}},  {\em JHEP} {\bf 1311}
  (2013) 191, [\href{http://xxx.lanl.gov/abs/1308.6114}{{\tt
  arXiv:1308.6114}}].

\bibitem{Braun:2014npa}
V.~Braun, S.~Derkachov, and A.~Manashov, {\it {Integrability of the evolution
  equations for heavy-light baryon distribution amplitudes}},
  \href{http://xxx.lanl.gov/abs/1406.0664}{{\tt arXiv:1406.0664}}.

\bibitem{Kruger:2005ep}
F.~Kr{\"u}ger and J.~Matias, {\it {Probing new physics via the transverse
  amplitudes of $B^0 \to K^{*0} (\to K^- \pi^+) \ell^+\ell^-$ at large
  recoil}},  {\em Phys.Rev.} {\bf D71} (2005) 094009,
  [\href{http://xxx.lanl.gov/abs/hep-ph/0502060}{{\tt hep-ph/0502060}}].

\bibitem{Altmannshofer:2008dz}
W.~Altmannshofer, P.~Ball, A.~Bharucha, A.~J. Buras, D.~M. Straub, et~al., {\it
  {Symmetries and Asymmetries of $B \to K^{*} \mu^{+} \mu^{-}$ Decays in the
  Standard Model and Beyond}},  {\em JHEP} {\bf 0901} (2009) 019,
  [\href{http://xxx.lanl.gov/abs/0811.1214}{{\tt arXiv:0811.1214}}].

\bibitem{Bobeth:2008ij}
C.~Bobeth, G.~Hiller, and G.~Piranishvili, {\it {CP Asymmetries in $\bar{B} \to
  \bar{K}^* (\to \bar{K} \pi) \bar{\ell} \ell$ and Untagged $\bar{B}_s$, $B_s
  \to \phi (\to K^{+} K^-) \bar{\ell} \ell$ Decays at NLO}},  {\em JHEP} {\bf
  0807} (2008) 106, [\href{http://xxx.lanl.gov/abs/0805.2525}{{\tt
  arXiv:0805.2525}}].

\bibitem{Egede:2008uy}
U.~Egede, T.~Hurth, J.~Matias, M.~Ramon, and W.~Reece, {\it {New observables in
  the decay mode $\bar{B}_{d} \to \bar{K}^{*0} \ell^+ \ell^-$}},  {\em JHEP}
  {\bf 0811} (2008) 032, [\href{http://xxx.lanl.gov/abs/0807.2589}{{\tt
  arXiv:0807.2589}}].

\bibitem{Bobeth:2010wg}
C.~Bobeth, G.~Hiller, and D.~van Dyk, {\it {The Benefits of $\bar{B} \to
  \bar{K}^* \ell^+ \ell^-$ Decays at Low Recoil}},  {\em JHEP} {\bf 1007}
  (2010) 098, [\href{http://xxx.lanl.gov/abs/1006.5013}{{\tt
  arXiv:1006.5013}}].

\bibitem{Becirevic:2011bp}
D.~Becirevic and E.~Schneider, {\it {On transverse asymmetries in $B \to K^*
  \ell^+\ell^-$}},  {\em Nucl.Phys.} {\bf B854} (2012) 321--339,
  [\href{http://xxx.lanl.gov/abs/1106.3283}{{\tt arXiv:1106.3283}}].

\bibitem{Bobeth:2012vn}
C.~Bobeth, G.~Hiller, and D.~van Dyk, {\it {General Analysis of $\bar{B} \to
  \bar{K}^{(*)}\ell^+ \ell^-$ Decays at Low Recoil}},  {\em Phys.Rev.} {\bf
  D87} (2013) 034016, [\href{http://xxx.lanl.gov/abs/1212.2321}{{\tt
  arXiv:1212.2321}}].

\bibitem{Mannel:1997xy}
T.~Mannel and S.~Recksiegel, {\it {Flavor changing neutral current decays of
  heavy baryons: The Case $\Lambda_b \to \Lambda \gamma$}},  {\em J.Phys.} {\bf
  G24} (1998) 979--990, [\href{http://xxx.lanl.gov/abs/hep-ph/9701399}{{\tt
  hep-ph/9701399}}].

\bibitem{Mannel:2011xg}
T.~Mannel and Y.-M. Wang, {\it {Heavy-to-light baryonic form factors at large
  recoil}},  {\em JHEP} {\bf 1112} (2011) 067,
  [\href{http://xxx.lanl.gov/abs/1111.1849}{{\tt arXiv:1111.1849}}].

\bibitem{Aaij:2013oxa}
{\bf LHCb collaboration} Collaboration, R.~Aaij et~al., {\it {Measurements of
  the $\Lambda_b^0 \to J/\psi \Lambda$ decay amplitudes and the $\Lambda_b^0$
  polarisation in $pp$ collisions at $\sqrt{s} = 7$ TeV}},  {\em Physics
  Letters B} {\bf 724} (2013) 27,
  [\href{http://xxx.lanl.gov/abs/1302.5578}{{\tt arXiv:1302.5578}}].

\bibitem{Buchalla:1995vs}
G.~Buchalla, A.~J. Buras, and M.~E. Lautenbacher, {\it {Weak decays beyond
  leading logarithms}},  {\em Rev.Mod.Phys.} {\bf 68} (1996) 1125--1144,
  [\href{http://xxx.lanl.gov/abs/hep-ph/9512380}{{\tt hep-ph/9512380}}].

\bibitem{Chetyrkin:1996vx}
K.~G. Chetyrkin, M.~Misiak, and M.~Munz, {\it {Weak radiative B meson decay
  beyond leading logarithms}},  {\em Phys.Lett.} {\bf B400} (1997) 206--219,
  [\href{http://xxx.lanl.gov/abs/hep-ph/9612313}{{\tt hep-ph/9612313}}].

\bibitem{Buras:2011we}
A.~J. Buras, {\it {Climbing NLO and NNLO Summits of Weak Decays}},
  \href{http://xxx.lanl.gov/abs/1102.5650}{{\tt arXiv:1102.5650}}.

\bibitem{Beneke:2001at}
M.~Beneke, T.~Feldmann, and D.~Seidel, {\it {Systematic approach to exclusive
  $B \to V \ell^+ \ell^-, V \gamma$ decays}},  {\em Nucl.Phys.} {\bf B612}
  (2001) 25--58, [\href{http://xxx.lanl.gov/abs/hep-ph/0106067}{{\tt
  hep-ph/0106067}}].

\bibitem{Beneke:2004dp}
M.~Beneke, T.~Feldmann, and D.~Seidel, {\it {Exclusive radiative and
  electroweak $b \to d$ and $b \to s$ penguin decays at NLO}},  {\em
  Eur.Phys.J.} {\bf C41} (2005) 173--188,
  [\href{http://xxx.lanl.gov/abs/hep-ph/0412400}{{\tt hep-ph/0412400}}].

\bibitem{Lee:1992ih}
C.~L. Lee, M.~Lu, and M.~B. Wise, {\it {$B_{l4}$ and $D_{l4}$ decay}},  {\em
  Phys.Rev.} {\bf D46} (1992) 5040--5048.

\bibitem{Faessler:2002ut}
A.~Faessler, T.~Gutsche, M.~Ivanov, J.~Korner, and V.~E. Lyubovitskij, {\it
  {The Exclusive rare decays $B \to$ K(K*) $\bar{\ell} \ell$ and $B_c \to$
  D(D*) $\bar{\ell} \ell$ in a relativistic quark model}},  {\em
  Eur.Phys.J.direct} {\bf C4} (2002) 18,
  [\href{http://xxx.lanl.gov/abs/hep-ph/0205287}{{\tt hep-ph/0205287}}].

\bibitem{Faller:2013dwa}
S.~Faller, T.~Feldmann, A.~Khodjamirian, T.~Mannel, and D.~van Dyk, {\it
  {Disentangling the Decay Observables in $B^- \to
  \pi^+\pi^-\ell^-\bar\nu_\ell$}},  {\em Phys.Rev.} {\bf D89} (2014) 014015,
  [\href{http://xxx.lanl.gov/abs/1310.6660}{{\tt arXiv:1310.6660}}].

\bibitem{Okun:1965}
L.~Okun, {\em Weak Interaction of Elementary Particles}.
\newblock 1965.

\bibitem{Hambrock:2012dg}
C.~Hambrock and G.~Hiller, {\it {Extracting $B \to K^*$ Form Factors from
  Data}},  {\em Phys.Rev.Lett.} {\bf 109} (2012) 091802,
  [\href{http://xxx.lanl.gov/abs/1204.4444}{{\tt arXiv:1204.4444}}].

\bibitem{Beaujean:2012uj}
F.~Beaujean, C.~Bobeth, D.~van Dyk, and C.~Wacker, {\it {Bayesian Fit of
  Exclusive $b \to s \bar\ell\ell$ Decays: The Standard Model Operator Basis}},
   {\em JHEP} {\bf 1208} (2012) 030,
  [\href{http://xxx.lanl.gov/abs/1205.1838}{{\tt arXiv:1205.1838}}].

\bibitem{Hambrock:2013zya}
C.~Hambrock, G.~Hiller, S.~Schacht, and R.~Zwicky, {\it {$B \to K^*$ Form
  Factors from Flavor Data to QCD and Back}},  {\em Phys.Rev.} {\bf D89} (2014)
  074014, [\href{http://xxx.lanl.gov/abs/1308.4379}{{\tt arXiv:1308.4379}}].

\bibitem{Beaujean:2013soa}
F.~Beaujean, C.~Bobeth, and D.~van Dyk, {\it {Comprehensive Bayesian Analysis
  of Rare (Semi)leptonic and Radiative B Decays}},
  \href{http://xxx.lanl.gov/abs/1310.2478}{{\tt arXiv:1310.2478}}.

\bibitem{Das:2014sra}
D.~Das, G.~Hiller, M.~Jung, and A.~Shires, {\it {The $ \overline{B}\to
  \overline{K}\pi \ell \ell $ and $ {\overline{B}}_s\ \to \overline{K}K\ell
  \ell $ distributions at low hadronic recoil}},  {\em JHEP} {\bf 1409} (2014)
  109, [\href{http://xxx.lanl.gov/abs/1406.6681}{{\tt arXiv:1406.6681}}].

\bibitem{Charles:1998dr}
J.~Charles, A.~Le~Yaouanc, L.~Oliver, O.~Pene, and J.~Raynal, {\it {Heavy to
  light form-factors in the heavy mass to large energy limit of QCD}},  {\em
  Phys.Rev.} {\bf D60} (1999) 014001,
  [\href{http://xxx.lanl.gov/abs/hep-ph/9812358}{{\tt hep-ph/9812358}}].

\bibitem{Beneke:2000wa}
M.~Beneke and T.~Feldmann, {\it {Symmetry breaking corrections to heavy to
  light B meson form-factors at large recoil}},  {\em Nucl.Phys.} {\bf B592}
  (2001) 3--34, [\href{http://xxx.lanl.gov/abs/hep-ph/0008255}{{\tt
  hep-ph/0008255}}].

\bibitem{Bauer:2000yr}
C.~W. Bauer, S.~Fleming, D.~Pirjol, and I.~W. Stewart, {\it {An Effective field
  theory for collinear and soft gluons: Heavy to light decays}},  {\em
  Phys.Rev.} {\bf D63} (2001) 114020,
  [\href{http://xxx.lanl.gov/abs/hep-ph/0011336}{{\tt hep-ph/0011336}}].

\bibitem{Beneke:2002ph}
M.~Beneke, A.~Chapovsky, M.~Diehl, and T.~Feldmann, {\it {Soft collinear
  effective theory and heavy to light currents beyond leading power}},  {\em
  Nucl.Phys.} {\bf B643} (2002) 431--476,
  [\href{http://xxx.lanl.gov/abs/hep-ph/0206152}{{\tt hep-ph/0206152}}].

\bibitem{Asatryan:2001zw}
H.~Asatryan, H.~Asatrian, C.~Greub, and M.~Walker, {\it {Calculation of two
  loop virtual corrections to $b \to s \ell^+ \ell^-$ in the standard model}},
  {\em Phys.Rev.} {\bf D65} (2002) 074004,
  [\href{http://xxx.lanl.gov/abs/hep-ph/0109140}{{\tt hep-ph/0109140}}].

\bibitem{EOS}
D.~van Dyk et~al., {\em {EOS: A HEP Program for Flavor Observables}}.
\newblock {The version used for this publication is available from
  \url{http://project.het.physik.tu-dortmund.de/source/eos/tag/lambdab}}.

\bibitem{Bona:2006ah}
{\bf UTfit Collaboration} Collaboration, M.~Bona et~al., {\it {The Unitarity
  Triangle Fit in the Standard Model and Hadronic Parameters from Lattice QCD:
  A Reappraisal after the Measurements of $\Delta m_s$ and BR($B \to \tau
  \nu_\tau$)}},  {\em JHEP} {\bf 0610} (2006) 081,
  [\href{http://xxx.lanl.gov/abs/hep-ph/0606167}{{\tt hep-ph/0606167}}]. We use
  the updated data from Summer 2013 (post-EPS13), as obtained from
  \url{http://www.utfit.org/UTfit/ResultsSummer2013PostEPS}.

\bibitem{Beringer:1900zz}
{\bf Particle Data Group} Collaboration, J.~Beringer et~al., {\it {Review of
  Particle Physics (RPP)}},  {\em Phys.Rev.} {\bf D86} (2012) 010001.

\bibitem{Amhis:2012bh}
{\bf Heavy Flavor Averaging Group} Collaboration, Y.~Amhis et~al., {\it
  {Averages of B-Hadron, C-Hadron, and tau-lepton properties as of early
  2012}},  \href{http://xxx.lanl.gov/abs/1207.1158}{{\tt arXiv:1207.1158}}. We
  use the update prepared for the PDG 2014 Review of Particle Physics.

\bibitem{Aaij:2013oha}
{\bf LHCb collaboration} Collaboration, R.~Aaij et~al., {\it {Precision
  measurement of the $\Lambda_b^0$ baryon lifetime}},  {\em Phys.Rev.Lett.}
  {\bf 111} (2013) 102003, [\href{http://xxx.lanl.gov/abs/1307.2476}{{\tt
  arXiv:1307.2476}}].

\bibitem{Aaij:2014owa}
{\bf LHCb collaboration} Collaboration, R.~Aaij et~al., {\it {Measurements of
  the $B^+, B^0, B^0_s$ meson and $\Lambda^0_b$ baryon lifetimes}},  {\em JHEP}
  {\bf 1404} (2014) 114, [\href{http://xxx.lanl.gov/abs/1402.2554}{{\tt
  arXiv:1402.2554}}].

\bibitem{Beylich:2011aq}
M.~Beylich, G.~Buchalla, and T.~Feldmann, {\it {Theory of $B \to
  K^{(*)}\ell^+\ell^-$ decays at high $q^2$: OPE and quark-hadron duality}},
  {\em Eur.Phys.J.} {\bf C71} (2011) 1635,
  [\href{http://xxx.lanl.gov/abs/1101.5118}{{\tt arXiv:1101.5118}}].

\bibitem{Aaij:2013pta}
{\bf LHCb collaboration} Collaboration, R.~Aaij et~al., {\it {Observation of a
  resonance in $B^+ \to K^+ \mu^+\mu^-$ decays at low recoil}},  {\em
  Phys.Rev.Lett.} {\bf 111} (2013), no.~11 112003,
  [\href{http://xxx.lanl.gov/abs/1307.7595}{{\tt arXiv:1307.7595}}].

\bibitem{Khodjamirian:2010vf}
A.~Khodjamirian, T.~Mannel, A.~Pivovarov, and Y.-M. Wang, {\it {Charm-loop
  effect in $B \to K^{(*)} \ell^{+} \ell^{-}$ and $B\to K^*\gamma$}},  {\em
  JHEP} {\bf 1009} (2010) 089, [\href{http://xxx.lanl.gov/abs/1006.4945}{{\tt
  arXiv:1006.4945}}].

\bibitem{Khodjamirian:2012rm}
A.~Khodjamirian, T.~Mannel, and Y.~Wang, {\it {$B \to K \ell^{+}\ell^{-}$ decay
  at large hadronic recoil}},  {\em JHEP} {\bf 1302} (2013) 010,
  [\href{http://xxx.lanl.gov/abs/1211.0234}{{\tt arXiv:1211.0234}}].

\bibitem{Lyon:2014hpa}
J.~Lyon and R.~Zwicky, {\it {Resonances gone topsy turvy - the charm of QCD or
  new physics in $b \to s \ell^+ \ell^-$?}},
  \href{http://xxx.lanl.gov/abs/1406.0566}{{\tt arXiv:1406.0566}}.

\bibitem{Aaij:2013qta}
{\bf LHCb collaboration} Collaboration, R.~Aaij et~al., {\it {Measurement of
  Form-Factor-Independent Observables in the Decay $B^{0} \to K^{*0} \mu^+
  \mu^-$}},  {\em Phys.Rev.Lett.} {\bf 111} (2013), no.~19 191801,
  [\href{http://xxx.lanl.gov/abs/1308.1707}{{\tt arXiv:1308.1707}}].

\bibitem{Jager:2012uw}
S.~J{\"a}ger and J.~Martin~Camalich, {\it {On $B \to V \ell^+ \ell^-$ at small
  dilepton invariant mass, power corrections, and new physics}},  {\em JHEP}
  {\bf 1305} (2013) 043, [\href{http://xxx.lanl.gov/abs/1212.2263}{{\tt
  arXiv:1212.2263}}].

\bibitem{Altmannshofer:2013foa}
W.~Altmannshofer and D.~M. Straub, {\it {New physics in $B \to K^*\mu\mu$?}},
  {\em Eur.Phys.J.} {\bf C73} (2013), no.~12 2646,
  [\href{http://xxx.lanl.gov/abs/1308.1501}{{\tt arXiv:1308.1501}}].

\bibitem{Descotes-Genon:2013wba}
S.~Descotes-Genon, J.~Matias, and J.~Virto, {\it {Understanding the $B \to
  K^*\mu^+\mu^-$ Anomaly}},  {\em Phys.Rev.} {\bf D88} (2013), no.~7 074002,
  [\href{http://xxx.lanl.gov/abs/1307.5683}{{\tt arXiv:1307.5683}}].

\bibitem{Grinstein:2004vb}
B.~Grinstein and D.~Pirjol, {\it {Exclusive rare $B \to K^* \ell^+ \ell^-$
  decays at low recoil: Controlling the long-distance effects}},  {\em
  Phys.Rev.} {\bf D70} (2004) 114005,
  [\href{http://xxx.lanl.gov/abs/hep-ph/0404250}{{\tt hep-ph/0404250}}].

\bibitem{Bourrely:2008za}
C.~Bourrely, I.~Caprini, and L.~Lellouch, {\it {Model-independent description
  of $B \to \pi \ell \bar\nu$ decays and a determination of $|V_{ub}|$}},  {\em
  Phys.Rev.} {\bf D79} (2009) 013008,
  [\href{http://xxx.lanl.gov/abs/0807.2722}{{\tt arXiv:0807.2722}}].

\bibitem{Khodjamirian:2011jp}
A.~Khodjamirian, C.~Klein, T.~Mannel, and Y.-M. Wang, {\it {Form Factors and
  Strong Couplings of Heavy Baryons from QCD Light-Cone Sum Rules}},  {\em
  JHEP} {\bf 1109} (2011) 106, [\href{http://xxx.lanl.gov/abs/1108.2971}{{\tt
  arXiv:1108.2971}}].

\bibitem{Jaynes:2003}
E.~Jaynes and G.~Bretthorst, {\em Probability Theory: The Logic of Science}.
\newblock Cambridge University Press, 2003.

\bibitem{Wang:2008sm}
Y.-M. Wang, Y.~Li, and C.-D. Lu, {\it {Rare Decays of $\Lambda_b \to
  \Lambda\gamma$ and $\Lambda_b \to \Lambda \ell^+ \ell^-$ in the Light-cone
  Sum Rules}},  {\em Eur.Phys.J.} {\bf C59} (2009) 861--882,
  [\href{http://xxx.lanl.gov/abs/0804.0648}{{\tt arXiv:0804.0648}}].

\bibitem{Wang:2009hra}
Y.-M. Wang, Y.-L. Shen, and C.-D. Lu, {\it {$\Lambda_b \to p, \Lambda$
  transition form factors from QCD light-cone sum rules}},  {\em Phys.Rev.}
  {\bf D80} (2009) 074012, [\href{http://xxx.lanl.gov/abs/0907.4008}{{\tt
  arXiv:0907.4008}}].

\bibitem{Haber:1994pe}
H.~E. Haber, {\it {Spin formalism and applications to new physics searches}},
  in {\em {21st Annual SLAC Summer Institute on Particle Physics: Spin
  structure in high energy processes}}, 1994.
\newblock \href{http://xxx.lanl.gov/abs/hep-ph/9405376}{{\tt hep-ph/9405376}}.

\end{thebibliography}\endgroup

\end{document}